\documentclass[fleqn,usenatbib]{mnras}

\usepackage{newtxtext,newtxmath,physics,comment}

\usepackage[T1]{fontenc}

\DeclareRobustCommand{\VAN}[3]{#2}
\let\VANthebibliography\thebibliography
\def\thebibliography{\DeclareRobustCommand{\VAN}[3]{##3}\VANthebibliography}

\usepackage{graphicx}	
\usepackage{amsmath}	
\usepackage{longtable}

\title[Massive RSG pulsation]{Radial pulsation runaway in massive red supergiants in late evolutionary stage and implications to hydrogen-rich supernovae}

\author[A. Suzuki and T. Shigeyama]{
Akihiro Suzuki,$^{1}$\thanks{E-mail: akihiro.suzuki@resceu.s.u-tokyo.ac.jp(AS)}
and Toshikazu Shigeyama$^{1}$
\\
$^{1}$Research Center for the Early Universe, The University of Tokyo, 7-3-1 Hongo, Bunkyo-ku, Tokyo 181-8588, Japan\\
}

\date{Accepted XXX. Received YYY; in original form ZZZ}

\pubyear{2025}

\begin{document}
\label{firstpage}
\pagerange{\pageref{firstpage}--\pageref{lastpage}}
\maketitle

\begin{abstract}
Performing a series of hydrodynamic stellar evolutionary simulations with \textsc{Mesa} (Module for Experiments in Stellar Astrophysics), we investigate the excitation and growth of radial pulsations of massive red supergiants (RSGs) with the initial mass range of $M_\mathrm{ini}=13$--$18\,\mathrm{M}_\odot$. 
We show that strong radial pulsations develop in the hydrogen-rich envelope in their late evolutionary stages, and eventually the surface radial velocity exceeds the escape velocity for higher-mass models. 
On the other hand, lower-mass models exhibit more moderate pulsations with finite velocity amplitudes and are expected to keep massive hydrogen-rich envelopes when they evolve toward the gravitational collapse of the iron core. 
While the latter group ends up as a familiar transient population of exploding RSGs, i.e., type IIP supernovae (SNe), the former group may expel a part of their envelopes and explode as different transients population. 
We investigate how the energy of the oscillating envelope is dissipated and released as radiation. 
We also empirically determine the condition for the pulsation-driven mass ejection in terms of the luminosity-to-mass ratio, $L/M>10^{3.9}\mathrm{L}_\odot/\mathrm{M}_\odot$. 
The corresponding luminosity threshold for the explored mass range may explain the observationally inferred constraints on type IIP SN progenitors. 
\end{abstract}

\begin{keywords}
stars: massive -- stars: oscillations -- stars: mass-loss -- supernovae: general
\end{keywords}



\section{Introduction}\label{sec:introduction} 
Massive stars are an essential driver of galaxy evolution through their mechanical and chemical feedback.  
Although the standard picture of massive star evolution toward gravitational collapse has long been established \citep[e.g.,][]{1978ApJ...225.1021W,1980SSRv...25..155S,1988PhR...163...13N,1995ApJS..101..181W,1996snih.book.....A,2002RvMP...74.1015W,2003ApJ...592..404L,2018ApJS..237...13L,2007PhR...442..269W,2012ARA&A..50..107L}, there are still a lot of questions remain unanswered. 
Especially, mapping massive stars with various initial masses to specific types of core-collapse supernovae (CCSNe) is still in extensive debate. 
One of the difficulties in connecting massive stars to CCSNe arises from the still unknown stellar activities and evolution at advanced nuclear burning stages. 
The massive star evolution after core helium depletion is governed by efficient neutrino cooling in the dense core and therefore rapidly proceeds. 
This short evolutionary time scale makes it difficult to find and scrutinize massive stars in such advanced burning stages. 
CCSN observations in both pre- and post-explosion stages instead offer a unique window to investigate massive stars' final activities. 

It is widely believed that most red supergiants (RSGs), if not all, explode in their final evolutionary stage and are observed as type IIP CCSNe (hereafter, SNe-IIP). 
However, even for this most commonly observed SN population, a discrepancy between the mass ranges of existing RSGs and SNe-IIP progenitors is suggested (the so-called RSG problem; \citealt{2009MNRAS.395.1409S,2009ARA&A..47...63S,2015PASA...32...16S,2020MNRAS.493.4945K}, see also \citealt{2020MNRAS.493..468D,2020MNRAS.496L.142D,2024arXiv241014027B}). 
Surveys of RSGs in our own and nearby galaxies have revealed RSGs as luminous as $L_\mathrm{bol}\simeq 10^{5.5}\mathrm{L}_\odot$, which corresponds to massive stars with the initial masses of $\simeq 25\,\mathrm{M}_\odot$ \citep[e.g.,][]{2003AJ....126.2867M,2005ApJ...628..973L,2006ApJ...645.1102L,2009ApJ...703..420M,2016ApJ...826..224M,2020ApJ...889...44N,2023ApJ...942...69M}. 
On the other hand, the progenitors of nearby SNe-IIP have been identified in archival images of space- and ground-based observatories \citep[e.g.,][]{2003PASP..115....1V,2003PASP..115.1289V,2004Sci...303..499S,2005MNRAS.364L..33M,2006ApJ...641.1060L,2017RSPTA.37560277V,2025arXiv250715973V}. 
In fact, they appear to be fainter than $L_\mathrm{bol}\simeq 10^{5.1}\mathrm{L}_\odot$ \citep{2009MNRAS.395.1409S,2009ARA&A..47...63S,2015PASA...32...16S,2022MNRAS.515..897R}. 
The apparent lack of SNe-IIP arising from most massive RSGs is also supported by systematic comparisons of SNe-IIP observations with synthetic light curves \citep{2022A&A...660A..40M,2022A&A...660A..41M,2022A&A...660A..42M} and nebular spectra \citep{2012A&A...546A..28J,2014MNRAS.439.3694J,2015MNRAS.448.2482J,2016MNRAS.459.3939V}. 
This discrepancy indicates that most massive RSGs may explode as transients other than SNe-IIP or might not explode at all. 

The latter idea leads to the so-called failed SN hypothesis; the gravitational collapse of an iron core fails to turn into a violent disruption of the star and instead most stellar mantle collapses to a black hole with a tiny fraction of the stellar envelope expelled \citep{1980Ap&SS..69..115N,2013ApJ...769..109L,2014ApJ...785...28K}. 
It is expected that such a collapse event results in a red transient and there are several failed SN candidates reported so far \citep{2021MNRAS.508..516N,2024arXiv241014778D}.  
However, one of the most promising failed SN candidates ever identified, N6946-BH1 \citep{2017MNRAS.468.4968A}, is now questioned by the recent James Webb Space Telescope observations (\citealt{2024ApJ...964..171B}, but, see also \citealt{2024ApJ...962..145K}), highlighting the difficulty of an unambiguous identification of a failed SN. 
In addition, first-principle simulations of the collapsing iron core by several independent groups show successful shock revival powered by neutrino heating for stars with the initial mass range of $\sim 10$--$25\,\mathrm{M}_\odot$ \citep{2021Natur.589...29B,2024ApJ...964L..16B,2024Ap&SS.369...80J,2024arXiv240508367N}, as opposed to the claimed mass intervals of non-explosion \citep[e.g.,][]{2012ApJ...757...69U,2016ApJ...818..124E,2016ApJ...821...38S}. 

With these recent findings, one may be intrigued by the alternative possibility; massive RSGs do explode, but ``disguise'' themselves right before their gravitational collapse \citep[e.g.,][]{2012MNRAS.419.2054W}. 
In other words, stars more massive than a certain threshold mass once evolve into RSGs after core H depletion, but they experience significant structural changes at some point on the course of RSG evolution, which place them away from the former locations in Hertzsprung-Russell (HR) diagram. 
The presence of massive RSGs in nearby galaxies suggest that their RSG lifetimes are somewhat long, thereby indicating that the disguise should happen quickly, e.g., after entering advanced nuclear burning stages. 
Indeed, it has been suggested that massive RSGs may experience enhanced mass-loss and turn into yellow supergiants/hypergiants \citep[e.g.,][]{1998A&ARv...8..145D,2012A&A...537A.146E,2012A&A...538L...8G}. 
However, the origin and properties of such enhanced mass-loss are not yet clarified. 

The pulsation-driven mass-loss is a promising mechanism to realize such brief and drastic structural changes in massive stars' envelope. 
The envelope of RSGs with high luminosity-to-mass ratio is known to be unstable to radial oscillations \citep[e.g.,][]{1994A&A...289..449L,1997A&A...327..224H,2002ApJ...565..559G}. 
\cite{1997A&A...327..224H} studied the pulsation properties of rotating $10$, $12$, $15$, and $20\,\mathrm{M}_\odot$ stars and find that radial pulsations more strongly grow for more massive stars at later evolutionary stages. 
They expect the occurrence of a ``superwind'' in the last several $10^4$ years before the core collapse. 
\cite{2010ApJ...717L..62Y} also investigated radial pulsations of RSGs, but for a wider initial mass range of $17$--$40\,\mathrm{M}_\odot$ and in several evolutionary stages. 
They indeed suggest that an enhanced mass-loss caused by the growing radial pulsation with supersonic surface velocities can explain the lack of massive SNe-IIP progenitors. 
The radial pulsation of a $25\,\mathrm{M}_\odot$ RSG is also demonstrated by \cite{2013ApJS..208....4P}, who developed and used the publicly available stellar hydrodynamics evolutionary code \textsc{Mesa} (Module for Experiments in Stellar Astrophysics; \citealt{2011ApJS..192....3P,2013ApJS..208....4P,2015ApJS..220...15P,2018ApJS..234...34P,2019ApJS..243...10P,2023ApJS..265...15J}). 
More recently, \cite{2018ApJ...853...79C} mention the appearance of supersonic layers in their \textsc{Mesa} RSG models with high luminosities, but without detailed analysis. 
\cite{clayton2018a} also used \textsc{Mesa} to directly simulate the pulsating envelopes of $15$--$40\,\mathrm{M}_\odot$ RSGs and explored the possibility of pulsation-driven dynamical mass-loss. 
Very recently\footnote{While this work was under peer review}, \cite{2025arXiv250804497S} have extended the pulsation-driven mass-loss prescription suggested by \cite{2010ApJ...717L..62Y} for RSG with an initial mass range of $12$--$20\,\mathrm{M}_\odot$. 
They argue that the enhanced mass-loss can reproduce CSM with properties consistent with those implied from several SNe-IIP observations. 
Despite these efforts, how the stellar envelope undergoes radial oscillations in the non-linear stage following the saturation of linear growth, and how these pulsations grow to large amplitudes and eventually lead to (possible) partial envelope ejection, remain open questions. 
It is also unclear whether a specific mass or luminosity threshold exists for pulsation-driven mass loss. 
Furthermore, to enable a more meaningful comparison between theoretical models and the growing number of detected SN progenitors, a denser stellar model grid around the proposed initial mass threshold would be beneficial.  

In this work, we conduct a series of evolutionary calculations of massive stars with \textsc{Mesa} and explore linear and non-linear radial pulsations in massive RSGs' envelope in the late evolutionary stage.
With its capability of evolutionary hydrodynamic simulations of stars, \textsc{Mesa} is now widely used for studies of massive RSGs, including their radial oscillations \citep[e.g.,][]{2020ApJ...902...63J}.
We show that models with high luminosity-to-mass ratios eventually experience pulsations with expansion velocities exceeding the surface escape velocity and then computations stop due to too small time steps. 
On the other hand, models with lower luminosity-to-mass ratios show regular pulsation cycles with surface velocity amplitudes below the escape velocity. 
We also clarify the response of the stellar envelopes to growing pulsations and empirically examine the condition for pulsation-driven mass-loss with our model grid. 
Then, we consider the final fates of massive RSGs.

This paper is organized in the following way. 
In Section 2, we describe our stellar evolutionary calculations. 
In Section 3, we take two models leading to different non-linear behaviours for examples and describe linear and non-linear evolution of radial pulsations. 
We then investigate the evolution of the stellar envelopes in the non-linear pulsation regime. 
In Section 4, we discuss the final fates of massive RSGs and observational implications. 
Finally, we conclude this paper in Section 5. 

\section{Stellar model grids}
In this section, we describe the setups and the summary of our stellar evolutionary simulations with \textsc{Mesa} (version 24.08.01). 
Our simulations consist of two model grids.
We first perform evolutionary simulations using the fiducial setups described below, continuing until iron core collapse. 
These 11 models (hereafter referred to as base models) cover an initial mass range of $13$–$18\,\mathrm{M}_\odot$. 
For each base model, we select eight epochs (or snapshots) covering evolutionary stages beyond central He depletion (approximately $0.5$–$20$ kyr before core collapse), from which we restart the simulations with significantly shorter time steps. 
Each restarted model is then evolved for an additional $100$ years to investigate the star’s linear and non-linear pulsation properties at each epoch. 
\begin{table*}
\centering
\caption{Base model grid and properties}
\label{table:model_description}
\begin{tabular}{rrrrrrrr}
\hline\hline
$M_\mathrm{ini}/\mathrm{M}_\odot$&
$M_\mathrm{presn}/\mathrm{M}_\odot$&
$M_\mathrm{He}/\mathrm{M}_\odot$&
$M_\mathrm{CO}/\mathrm{M}_\odot$&
$R_\mathrm{presn}/R_\odot$&
$\log_{10}(L_\mathrm{presn}/\mathrm{L}_\odot)$&
$\log_{10}(t_\mathrm{cc}/\mathrm{yr})$
\\
\hline 
$13.0$&$11.90$&$4.08$&$2.52$&$710$&$4.92$&$7.24$\\
$13.5$&$12.31$&$4.29$&$2.68$&$738$&$4.96$&$7.21$\\
$14.0$&$12.78$&$4.42$&$2.77$&$749$&$4.98$&$7.19$\\
$14.5$&$12.82$&$4.50$&$2.85$&$765$&$4.99$&$7.17$\\
$15.0$&$12.31$&$4.93$&$3.16$&$834$&$5.05$&$7.15$\\
$15.5$&$13.05$&$5.05$&$3.28$&$852$&$5.07$&$7.13$\\
$16.0$&$12.18$&$5.54$&$3.65$&$921$&$4.92$&$7.11$\\
$16.5$&$11.86$&$5.85$&$3.91$&$996$&$5.16$&$7.09$\\
$17.0$&$13.30$&$5.74$&$3.84$&$940$&$5.14$&$7.07$\\
$17.5$&$12.44$&$6.05$&$4.08$&$977$&$5.17$&$7.06$\\
$18.0$&$12.46$&$6.30$&$4.30$&$1032$&$5.20$&$7.04$\\
\hline\hline
\end{tabular}
\end{table*}
  
\begin{table*}
\centering
\caption{Short-timestep models with $M_\mathrm{ini}=13$--$15\,\mathrm{M}_\odot$ and properties}
\label{table:model_description2}
\begin{tabular}{lrrrrrrrrrr}
\hline\hline
model&
$M_\mathrm{ini}/\mathrm{M}_\odot$&
$N_\mathrm{zone}$&
$t_\mathrm{cc}-t_\mathrm{ini}[\mathrm{kyr}]$&
$M/\mathrm{M}_\odot$&
$\log_{10}(L/\mathrm{L}_\odot)$&
$R/R_\odot$&
$P_\mathrm{gyre}[\mathrm{yr}]$&
$P_\mathrm{lin}[\mathrm{yr}]$&
final state&
\\
\hline 
\texttt{M0130\_model1}&$13.0$&$ 3429 $&$ 20.3 $&$ 11.95 $&$ 4.73 $&$ 510 $&$ 0.64 $&$ 0.64 $& limit cycle\\
\texttt{M0130\_model2}&$13.0$&$ 3253 $&$ 10.4 $&$ 11.93 $&$ 4.84 $&$ 620 $&$ 0.92 $&$ 0.93 $& limit cycle\\
\texttt{M0130\_model3}&$13.0$&$ 3209 $&$ 7.79 $&$ 11.92 $&$ 4.87 $&$ 643 $&$ 0.99 $&$ 0.99 $& limit cycle\\
\texttt{M0130\_model4}&$13.0$&$ 3246 $&$ 5.86 $&$ 11.92 $&$ 4.88 $&$ 659 $&$ 1.03 $&$ 1.04 $& limit cycle\\
\texttt{M0130\_model5}&$13.0$&$ 3330 $&$ 4.07 $&$ 11.91 $&$ 4.89 $&$ 673 $&$ 1.07 $&$ 1.08 $& limit cycle\\
\texttt{M0130\_model6}&$13.0$&$ 3368 $&$ 2.08 $&$ 11.91 $&$ 4.90 $&$ 685 $&$ 1.11 $&$ 1.13 $& limit cycle\\
\texttt{M0130\_model7}&$13.0$&$ 3417 $&$ 0.94 $&$ 11.90 $&$ 4.92 $&$ 700 $&$ 1.16 $&$ 1.17 $& limit cycle\\
\texttt{M0130\_model8}&$13.0$&$ 3553 $&$ 0.52 $&$ 11.90 $&$ 4.92 $&$ 707 $&$ 1.19 $&$ 1.19 $& limit cycle\\
\hline
\texttt{M0135\_model1}&$13.5$&$ 3497 $&$ 20.1 $&$ 12.37 $&$ 4.73 $&$ 501 $&$ 0.61 $&$ 0.61 $& limit cycle\\
\texttt{M0135\_model2}&$13.5$&$ 3348 $&$ 10.3 $&$ 12.35 $&$ 4.87 $&$ 639 $&$ 0.95 $&$ 0.96 $& limit cycle\\
\texttt{M0135\_model3}&$13.5$&$ 3312 $&$ 8.10 $&$ 12.35 $&$ 4.89 $&$ 662 $&$ 1.02 $&$ 1.03 $& limit cycle\\
\texttt{M0135\_model4}&$13.5$&$ 3257 $&$ 5.94 $&$ 12.34 $&$ 4.91 $&$ 684 $&$ 1.08 $&$ 1.09 $& limit cycle\\
\texttt{M0135\_model5}&$13.5$&$ 3316 $&$ 3.93 $&$ 12.33 $&$ 4.93 $&$ 702 $&$ 1.14 $&$ 1.15 $& limit cycle\\
\texttt{M0135\_model6}&$13.5$&$ 3391 $&$ 1.85 $&$ 12.32 $&$ 4.94 $&$ 716 $&$ 1.19 $&$ 1.20 $& limit cycle\\
\texttt{M0135\_model7}&$13.5$&$ 3427 $&$ 1.04 $&$ 12.32 $&$ 4.95 $&$ 725 $&$ 1.22 $&$ 1.23 $& limit cycle\\
\texttt{M0135\_model8}&$13.5$&$ 3521 $&$ 0.50 $&$ 12.32 $&$ 4.95 $&$ 734 $&$ 1.25 $&$ 1.26 $& limit cycle\\
\hline
\texttt{M0140\_model1}&$14.0$&$ 3516 $&$ 20.1 $&$ 12.84 $&$ 4.74 $&$ 501 $&$ 0.59 $&$ 0.60 $& limit cycle\\
\texttt{M0140\_model2}&$14.0$&$ 3383 $&$ 10.4 $&$ 12.81 $&$ 4.88 $&$ 643 $&$ 0.94 $&$ 0.95 $& limit cycle\\
\texttt{M0140\_model3}&$14.0$&$ 3324 $&$ 8.11 $&$ 12.81 $&$ 4.91 $&$ 669 $&$ 1.02 $&$ 1.02 $& limit cycle\\
\texttt{M0140\_model4}&$14.0$&$ 3312 $&$ 5.87 $&$ 12.80 $&$ 4.93 $&$ 693 $&$ 1.09 $&$ 1.10 $& limit cycle\\
\texttt{M0140\_model5}&$14.0$&$ 3340 $&$ 4.03 $&$ 12.79 $&$ 4.95 $&$ 711 $&$ 1.15 $&$ 1.15 $& limit cycle\\
\texttt{M0140\_model6}&$14.0$&$ 3432 $&$ 1.97 $&$ 12.78 $&$ 4.96 $&$ 727 $&$ 1.20 $&$ 1.21 $& limit cycle\\
\texttt{M0140\_model7}&$14.0$&$ 3424 $&$ 1.14 $&$ 12.78 $&$ 4.96 $&$ 734 $&$ 1.20 $&$ 1.23 $& limit cycle\\
\texttt{M0140\_model8}&$14.0$&$ 3471 $&$ 0.51 $&$ 12.78 $&$ 4.97 $&$ 743 $&$ 1.22 $&$ 1.26 $& limit cycle\\
\hline
\texttt{M0145\_model1}&$14.5$&$ 3669 $&$ 20.5 $&$ 12.89 $&$ 4.81 $&$ 567 $&$ 0.74 $&$ 0.74 $& limit cycle\\
\texttt{M0145\_model2}&$14.5$&$ 3464 $&$ 10.2 $&$ 12.86 $&$ 4.90 $&$ 656 $&$ 0.98 $&$ 0.98 $& limit cycle\\
\texttt{M0145\_model3}&$14.5$&$ 3458 $&$ 8.00 $&$ 12.85 $&$ 4.92 $&$ 682 $&$ 1.05 $&$ 1.06 $& limit cycle\\
\texttt{M0145\_model4}&$14.5$&$ 3386 $&$ 5.84 $&$ 12.84 $&$ 4.94 $&$ 703 $&$ 1.12 $&$ 1.13 $& limit cycle\\
\texttt{M0145\_model5}&$14.5$&$ 3339 $&$ 4.05 $&$ 12.84 $&$ 4.96 $&$ 722 $&$ 1.18 $&$ 1.19 $& limit cycle\\
\texttt{M0145\_model6}&$14.5$&$ 3370 $&$ 1.85 $&$ 12.83 $&$ 4.97 $&$ 741 $&$ 1.24 $&$ 1.25 $& limit cycle\\
\texttt{M0145\_model7}&$14.5$&$ 3386 $&$ 1.05 $&$ 12.82 $&$ 4.98 $&$ 749 $&$ 1.27 $&$ 1.27 $& limit cycle\\
\texttt{M0145\_model8}&$14.5$&$ 3431 $&$ 0.50 $&$ 12.82 $&$ 4.99 $&$ 760 $&$ 1.30 $&$ 1.32 $& limit cycle\\
\hline
\texttt{M0150\_model1}&$15.0$&$ 3690 $&$ 20.2 $&$ 12.41 $&$ 4.96 $&$ 719 $&$ 1.19 $&$ 1.18 $& limit cycle\\
\texttt{M0150\_model2}&$15.0$&$ 3630 $&$ 10.5 $&$ 12.37 $&$ 4.94 $&$ 699 $&$ 1.13 $&$ 1.13 $& limit cycle\\
\texttt{M0150\_model3}&$15.0$&$ 3508 $&$ 8.15 $&$ 12.36 $&$ 4.96 $&$ 722 $&$ 1.20 $&$ 1.18 $& limit cycle\\
\texttt{M0150\_model4}&$15.0$&$ 3439 $&$ 6.18 $&$ 12.35 $&$ 4.98 $&$ 752 $&$ 1.30 $&$ 1.27 $& limit cycle\\
\texttt{M0150\_model5}&$15.0$&$ 3413 $&$ 3.97 $&$ 12.34 $&$ 5.01 $&$ 788 $&$ 1.43 $&$ 1.40 $& limit cycle\\
\texttt{M0150\_model6}&$15.0$&$ 3406 $&$ 2.10 $&$ 12.33 $&$ 5.03 $&$ 812 $&$ 1.52 $&$ 1.55 $& runaway\\
\texttt{M0150\_model7}&$15.0$&$ 3401 $&$ 0.97 $&$ 12.32 $&$ 5.04 $&$ 824 $&$ 1.57 $&$ 1.59 $& runaway\\
\texttt{M0150\_model8}&$15.0$&$ 3446 $&$ 0.47 $&$ 12.32 $&$ 5.04 $&$ 830 $&$ 1.59 $&$ 1.61 $& limit cycle\\
\hline\hline
\end{tabular}
\end{table*}

\begin{table*}
\centering
\caption{Short-timestep models with $M_\mathrm{ini}=15.5$--$18\,\mathrm{M}_\odot$ and properties}
\label{table:model_description3}
\begin{tabular}{lrrrrrrrrrr}
\hline\hline
model&
$M_\mathrm{ini}/\mathrm{M}_\odot$&
$N_\mathrm{zone}$&
$t_\mathrm{cc}-t_\mathrm{ini}[\mathrm{kyr}]$&
$M/\mathrm{M}_\odot$&
$\log_{10}(L/\mathrm{L}_\odot)$&
$R/R_\odot$&
$P_\mathrm{gyre}[\mathrm{yr}]$&
$P_\mathrm{lin}[\mathrm{yr}]$&
final state&
\\
\hline 
\texttt{M0155\_model1}&$15.5$&$ 3712 $&$ 20.0 $&$ 13.15 $&$ 4.95 $&$ 704 $&$ 1.10 $&$ 1.10 $& limit cycle\\
\texttt{M0155\_model2}&$15.5$&$ 3654 $&$ 10.3 $&$ 13.11 $&$ 4.96 $&$ 707 $&$ 1.11 $&$ 1.12 $& limit cycle\\
\texttt{M0155\_model3}&$15.5$&$ 3580 $&$ 8.00 $&$ 13.10 $&$ 4.98 $&$ 739 $&$ 1.21 $&$ 1.22 $& limit cycle\\
\texttt{M0155\_model4}&$15.5$&$ 3516 $&$ 6.05 $&$ 13.09 $&$ 5.00 $&$ 767 $&$ 1.30 $&$ 1.30 $& limit cycle\\
\texttt{M0155\_model5}&$15.5$&$ 3497 $&$ 3.88 $&$ 13.08 $&$ 5.03 $&$ 801 $&$ 1.42 $&$ 1.43 $& limit cycle\\
\texttt{M0155\_model6}&$15.5$&$ 3495 $&$ 2.04 $&$ 13.06 $&$ 5.05 $&$ 823 $&$ 1.50 $&$ 1.52 $& runaway\\
\texttt{M0155\_model7}&$15.5$&$ 3427 $&$ 0.93 $&$ 13.06 $&$ 5.06 $&$ 836 $&$ 1.55 $&$ 1.57 $& runaway\\
\texttt{M0155\_model8}&$15.5$&$ 3450 $&$ 0.54 $&$ 13.06 $&$ 5.06 $&$ 838 $&$ 1.56 $&$ 1.59 $& runaway\\
\hline
\texttt{M0160\_model1}&$16.0$&$ 4457 $&$ 20.3 $&$ 12.31 $&$ 5.00 $&$ 767 $&$ 1.35 $&$ 1.36 $& runaway\\
\texttt{M0160\_model2}&$16.0$&$ 3774 $&$ 10.5 $&$ 12.25 $&$ 5.04 $&$ 810 $&$ 1.52 $&$ 1.52 $& runaway\\
\texttt{M0160\_model3}&$16.0$&$ 3721 $&$ 8.15 $&$ 12.24 $&$ 5.04 $&$ 809 $&$ 1.51 $&$ 1.52 $& runaway\\
\texttt{M0160\_model4}&$16.0$&$ 3600 $&$ 6.12 $&$ 12.23 $&$ 5.05 $&$ 825 $&$ 1.58 $&$ 1.54 $& runaway\\
\texttt{M0160\_model5}&$16.0$&$ 3566 $&$ 4.09 $&$ 12.21 $&$ 5.07 $&$ 865 $&$ 1.74 $&$ 1.68 $& runaway\\
\texttt{M0160\_model6}&$16.0$&$ 3558 $&$ 2.04 $&$ 12.20 $&$ 5.10 $&$ 906 $&$ 1.92 $&$ 1.87 $& runaway\\
\texttt{M0160\_model7}&$16.0$&$ 3610 $&$ 1.05 $&$ 12.19 $&$ 5.11 $&$ 920 $&$ 1.99 $&$ 1.99 $& runaway\\
\texttt{M0160\_model8}&$16.0$&$ 3572 $&$ 0.56 $&$ 12.19 $&$ 5.12 $&$ 929 $&$ 2.03 $&$ 2.02 $& runaway\\
\hline
\texttt{M0165\_model1}&$16.5$&$ 4265 $&$ 20.2 $&$ 12.01 $&$ 5.04 $&$ 807 $&$ 1.53 $&$ 1.51 $& runaway\\
\texttt{M0165\_model2}&$16.5$&$ 3757 $&$ 10.4 $&$ 11.94 $&$ 5.09 $&$ 873 $&$ 1.82 $&$ 1.81 $& runaway\\
\texttt{M0165\_model3}&$16.5$&$ 3762 $&$ 8.14 $&$ 11.92 $&$ 5.08 $&$ 860 $&$ 1.76 $&$ 1.75 $& runaway\\
\texttt{M0165\_model4}&$16.5$&$ 3662 $&$ 6.09 $&$ 11.91 $&$ 5.08 $&$ 866 $&$ 1.79 $&$ 1.77 $& runaway\\
\texttt{M0165\_model5}&$16.5$&$ 3588 $&$ 4.02 $&$ 11.89 $&$ 5.11 $&$ 905 $&$ 1.96 $&$ 1.89 $& runaway\\
\texttt{M0165\_model6}&$16.5$&$ 3571 $&$ 2.09 $&$ 11.88 $&$ 5.14 $&$ 947 $&$ 2.17 $&$ 2.12 $& runaway\\
\texttt{M0165\_model7}&$16.5$&$ 3621 $&$ 0.99 $&$ 11.86 $&$ 5.15 $&$ 966 $&$ 2.27 $&$ 2.22 $& runaway\\
\texttt{M0165\_model8}&$16.5$&$ 3673 $&$ 0.50 $&$ 11.86 $&$ 5.16 $&$ 978 $&$ 2.33 $&$ 2.26 $& runaway\\
\hline
\texttt{M0170\_model1}&$17.0$&$ 4229 $&$ 20.8 $&$ 13.43 $&$ 5.00 $&$ 733 $&$ 1.17 $&$ 1.18 $& limit cycle\\
\texttt{M0170\_model2}&$17.0$&$ 3687 $&$ 10.3 $&$ 13.37 $&$ 5.05 $&$ 796 $&$ 1.39 $&$ 1.39 $& limit cycle\\
\texttt{M0170\_model3}&$17.0$&$ 3641 $&$ 8.00 $&$ 13.36 $&$ 5.04 $&$ 793 $&$ 1.38 $&$ 1.38 $& runaway\\
\texttt{M0170\_model4}&$17.0$&$ 3578 $&$ 6.00 $&$ 13.34 $&$ 5.06 $&$ 823 $&$ 1.49 $&$ 1.46 $& runaway\\
\texttt{M0170\_model5}&$17.0$&$ 3489 $&$ 3.99 $&$ 13.33 $&$ 5.10 $&$ 876 $&$ 1.69 $&$ 1.67 $& runaway\\
\texttt{M0170\_model6}&$17.0$&$ 3536 $&$ 1.96 $&$ 13.31 $&$ 5.13 $&$ 917 $&$ 1.86 $&$ 1.87 $& runaway\\
\texttt{M0170\_model7}&$17.0$&$ 3581 $&$ 1.00 $&$ 13.30 $&$ 5.14 $&$ 935 $&$ 1.94 $&$ 1.94 $& runaway\\
\texttt{M0170\_model8}&$17.0$&$ 3572 $&$ 0.57 $&$ 13.30 $&$ 5.14 $&$ 940 $&$ 1.96 $&$ 1.96 $& runaway\\
\hline
\texttt{M0175\_model1}&$17.5$&$ 4230 $&$ 20.4 $&$ 12.61 $&$ 5.07 $&$ 828 $&$ 1.58 $&$ 1.58 $& runaway\\
\texttt{M0175\_model2}&$17.5$&$ 3731 $&$ 10.2 $&$ 12.53 $&$ 5.12 $&$ 900 $&$ 1.89 $&$ 1.83 $& runaway\\
\texttt{M0175\_model3}&$17.5$&$ 3798 $&$ 7.96 $&$ 12.51 $&$ 5.11 $&$ 879 $&$ 1.81 $&$ 1.81 $& runaway\\
\texttt{M0175\_model4}&$17.5$&$ 3743 $&$ 5.95 $&$ 12.50 $&$ 5.11 $&$ 874 $&$ 1.79 $&$ 1.76 $& runaway\\
\texttt{M0175\_model5}&$17.5$&$ 3644 $&$ 4.25 $&$ 12.48 $&$ 5.12 $&$ 893 $&$ 1.87 $&$ 1.80 $& runaway\\
\texttt{M0175\_model6}&$17.5$&$ 3603 $&$ 1.98 $&$ 12.46 $&$ 5.16 $&$ 951 $&$ 2.14 $&$ 2.09 $& runaway\\
\texttt{M0175\_model7}&$17.5$&$ 3650 $&$ 0.99 $&$ 12.45 $&$ 5.17 $&$ 972 $&$ 2.25 $&$ 2.23 $& runaway\\
\texttt{M0175\_model8}&$17.5$&$ 3710 $&$ 0.51 $&$ 12.45 $&$ 5.18 $&$ 984 $&$ 2.31 $&$ 2.30 $& runaway\\
\hline
\texttt{M0180\_model1}&$18.0$&$ 4170 $&$ 20.8 $&$ 12.66 $&$ 5.11 $&$ 870 $&$ 1.74 $&$ 1.72 $& runaway\\
\texttt{M0180\_model2}&$18.0$&$ 3691 $&$ 10.1 $&$ 12.57 $&$ 5.16 $&$ 957 $&$ 2.14 $&$ 2.04 $& runaway\\
\texttt{M0180\_model3}&$18.0$&$ 3801 $&$ 7.86 $&$ 12.54 $&$ 5.15 $&$ 932 $&$ 2.03 $&$ 2.02 $& runaway\\
\texttt{M0180\_model4}&$18.0$&$ 3777 $&$ 5.86 $&$ 12.52 $&$ 5.14 $&$ 921 $&$ 1.99 $&$ 2.00 $& runaway\\
\texttt{M0180\_model5}&$18.0$&$ 3675 $&$ 4.17 $&$ 12.51 $&$ 5.15 $&$ 930 $&$ 2.03 $&$ 1.94 $& runaway\\
\texttt{M0180\_model6}&$18.0$&$ 3618 $&$ 2.10 $&$ 12.49 $&$ 5.18 $&$ 976 $&$ 2.26 $&$ 2.18 $& runaway\\
\texttt{M0180\_model7}&$18.0$&$ 3661 $&$ 1.02 $&$ 12.48 $&$ 5.19 $&$ 1007 $&$ 2.42 $&$ 2.32 $& runaway\\
\texttt{M0180\_model8}&$18.0$&$ 3737 $&$ 0.49 $&$ 12.47 $&$ 5.20 $&$ 1017 $&$ 2.48 $&$ 2.40 $& runaway\\
\hline\hline
\end{tabular}
\end{table*}

\subsection{Base models}
In the following, we describe the setups of the base models. 
The properties of the models are summarized in Table \ref{table:model_description}. 

\subsubsection{Initial conditions, composition, and atmosphere}
We carry out simulations of non-rotating stars. 
The base model series contains $11$ models with the initial mass increased by $0.5\,\mathrm{M}_\odot$ from $13\,\mathrm{M}_\odot$ to $18\,\mathrm{M}_\odot$. 
The computations employ \texttt{20M\_pre\_ms\_to\_core\_collapse} inlists, which are provided as one of the \textsc{Mesa} test suites, with some modifications described below. 

For all models, the initial hydrogen, helium, and metal mass fractions are set to $(X_\mathrm{H},X_\mathrm{He},X_\mathrm{Z})=(0.7155,0.2703,0.0142)$, i.e., solar values are assumed. 
We employ a simplified nuclear reaction network composed of 21 nuclei (\texttt{approx21\_cr60\_plus\_co56.net}) with the widely used solar abundance in \cite{2009ARA&A..47..481A} as the initial chemical composition. 
The opacity table is accordingly selected.

According to the default setup, we set the parameter \texttt{tau\_factor}, which controls the outer boundary, equal to unity. 
With this value, the code solves the atmosphere of the star up to the optical depth of $\tau=2/3$, beyond which some temperature-optical depth relation is employed. 
We then assume an atmospheric treatment enabled by the following options; \texttt{atm\_option='T\_tau'}, \texttt{atm\_T\_tau\_relation = 'Eddington'}, and \texttt{atm\_T\_tau\_opacity = 'fixed'}.

\subsubsection{Time step and mesh parameters}
\textsc{Mesa} employs non-uniform time steps, which adaptively change in different evolutionary stages. 
The code provides some adjustable parameters for the time step control. 
Up to the end of carbon burning stage, we impose the following restrictions;
\begin{itemize}
    \item \texttt{delta\_lgRho\_cntr\_limit = 1.0d-3}
    \item \texttt{delta\_lgRho\_cntr\_hard\_limit = 1.0d-3}
    \item \texttt{delta\_lgRho\_limit = 1.0d-3}
    \item \texttt{delta\_lgT\_cntr\_limit = 1.0d-3}
    \item \texttt{delta\_lgT\_cntr\_hard\_limit = 1.0d-3}
    \item \texttt{delta\_lgT\_max\_limit = 1.0d-3}
    \item \texttt{delta\_lgT\_max\_hard\_limit = 1.0d-3}
    \item \texttt{dX\_nuc\_drop\_limit = 1.0d-3}
    \item \texttt{dX\_nuc\_drop\_min\_X\_limit = 1.0d-3}.
\end{itemize}
These parameters restrict the maximum change in several physical variables (central density, central/maximum temperature, mass fraction of nuclei, and so on) in a single time step. 

The computations are continued until the central carbon mass fraction drops below $10^{-5}$. 
We then continue computations toward the core collapse, but with relaxed time step restrictions as follows;
\begin{itemize}
    \item \texttt{delta\_lgRho\_cntr\_limit = 1.0d-2}
    \item \texttt{delta\_lgRho\_cntr\_hard\_limit = 1.0d-2}
    \item \texttt{delta\_lgRho\_limit = 1.0d-2}
    \item \texttt{dX\_nuc\_drop\_limit = 1.0d-2}.
\end{itemize}

For simulation mass grid, we set the parameter \texttt{delta\_mesh\_coefficient=0.5}. 
With this value, the star is typically covered by $\simeq 3200$--$4500$ mass zones in the carbon burning stage. 

\subsubsection{Mixing}
Mixing of materials in convective regions in a star is essential to determine the efficiency of the energy transport and the core growth. 
Unfortunately, there is no parameter-free treatment for convective mixing. 
\textsc{Mesa} offers some options to treat convection and convective core-boundary mixing. 
In \textsc{Mesa}, the treatment of convective mixing is based on the mixing-length theory (MLT) (\citealt{2018ApJS..234...34P,2019ApJS..243...10P}; see also \citealt{1999A&A...346..111L}). 
We adopt the Ledoux criterion with the mixing-length parameter fixed to be $\alpha_\mathrm{mlt}=2.5$ throughout the simulations. 
According to the default setups of \texttt{20M\_pre\_ms\_to\_core\_collapse} inlists, we utilize the so-called time-dependent convection (\texttt{TDC}) option. 
The radiative damping parameter $\alpha_\mathrm{r}$ (\texttt{alpha\_TDC\_DAMPR}) is set to zero. 
Other controlling parameters are also set to default values; $\alpha=2$, $\alpha_\mathrm{D}=1$, and $\alpha_{P_t}=0$.  
We note that \texttt{TDC} option with a default parameter set in \textsc{Mesa} is designed so that the treatment reduces to the MLT treatment described in \cite{1968pss..book.....C} for long time step values (Sec. 3 of \citealt{2023ApJS..265...15J} and references therein). 
In RSGs with extensive convective envelopes, several effects not included in this simplified setup, such as radiative damping of convective velocities, could play a significant role. 
Therefore, while we adopt this default option, it is not necessarily superior to alternative choices. Further explorations of convection parameters and their impact on outcomes is warranted. 
A semi-convection parameter of $\alpha_\mathrm{sc}=0.01$ is adopted. 
Neither \texttt{MLT++} nor \texttt{superadiabatic\ reduction} options are enabled.  
For the dependence of RSG evolution on the mixing parameters and code implementation, we refer the readers to the work by \cite{2018ApJ...853...79C}. 

The core-boundary mixing is of particular importance in determining the initial mass-core mass relation (e.g., \citealt{1987A&A...182..243M,2019A&A...622A..50H,2024A&A...682A.123T}). 
The overshooting parameter is often used to specify the mixed layer around the core-envelope interface. 
The overshooting distance is parameterized by a fraction of the pressure scale height, $d_\mathrm{ov}=f_\mathrm{ov}H_p$, where $f_\mathrm{ov}$ is a numerical parameter. 
In practice, \texttt{overshoot\_f(1) = 0.21} and \texttt{overshoot\_f0(1) = 0.01} are chosen, where the former and latter parameters specify the lengths of the mixed layers above and below the core-envelope interface. 


\subsubsection{Mass-loss}\label{sec:mass_loss}
We employ the default mass-loss formula used in the massive star test suite, i.e., the so-called Dutch scheme. 
It is a combination of the three mass-loss prescriptions based on observational results for OB stars by \cite{2001A&A...369..574V} for high effective temperature, O-M stars by \cite{1988A&AS...72..259D} for low effective temperature, and Wolf-Rayet stars by \cite{2000A&A...360..227N} (when surface hydrogen mass fraction drops to $X_\mathrm{H}<0.4$).  
The threshold temperatures connecting the former two prescriptions are set to $T_\mathrm{high}=1.2\times 10^4\,\mathrm{K}$ and $T_\mathrm{low}=8\times 10^3\,\mathrm{K}$, between which the mass-loss rate is determined by a linear interpolation. 
For all the models, we assume the wind efficiency parameter of $\eta_\mathrm{wind}=1.0$. 

\subsubsection{Hydrodynamics solver}
In this study, we utilize the hydrodynamics solver based on artificial viscosity implemented in \textsc{Mesa} (\texttt{v\_flag=.true.}). 
This allows us to investigate non-linear pulsating behaviours of RSGs after the linear growth is saturated. 
We do not utilize the hydrodynamic drag options. 

\textsc{Mesa} has the so-called \texttt{RSP} module \citep{2019ApJS..243...10P} based on \cite{2008AcA....58..193S}, which is dedicated for simulating non-linear pulsations of variable stars. 
While it has successfully demonstrated non-linear pulsations of low-mass variables, such as RR Lyrae \citep{2019ApJS..243...10P}, treating pulsating stars with large luminosity-to-mass ratios has some difficulties (see also \citealt{2020ApJ...902...63J}). 
Therefore, we opt to use the hydrodynamic solver for our simulations. 

A remark on the use of the \textsc{Mesa} hydrodynamics solver for pulsation studies is warranted.
The hydrodynamics module has been employed to study the pulsational behaviour of massive stars in several previous works \citep[e.g.,][]{2013ApJS..208....4P,2020ApJ...902...63J}.
These studies have demonstrated that the simulations can reproduce pulsation periods consistent with those predicted by linear perturbation analysis.
However, the amplitude, growth, and saturation of the pulsations are more complex functions of the parameters governing convection and the choice of numerical viscosity. 
We therefore caution that these parameters, which can significantly influence the pulsational properties, will require careful calibration against observations in future studies. 

\subsection{Short-timestep models}
\subsubsection{Initial models}

The initial models for this simulation series are taken from the base models described above. 
For each base model with a given initial mass, we extract eight snapshots to serve as initial conditions for the short-timestep model grid. 
These snapshots are chosen such that the time remaining until core collapse approximately corresponds to $t_\mathrm{cc} - t \simeq 20$, $10$, $8$, $6$, $4$, $2$, $1$, and $0.5 \times 10^3\,\mathrm{yr}$ (see Tables \ref{table:model_description2} and \ref{table:model_description3}). 

\subsubsection{Time steps and mesh parameters}
We restrict the time step in these simulations to significantly smaller values than those used in the base models. 
Specifically, we set the maximum time step to $\Delta t_\mathrm{max} = 2.5 \times 10^{-4}\,\mathrm{yr}$ ($< 0.1\,\mathrm{days}$), which is much shorter than the expected radial pulsation period ($1$–$2$ years) of RSGs in the late evolutionary stage. 
This restriction enables the simulations to resolve the physical processes that lead to the saturation of linearly growing pulsations and the dissipation of pulsation energy during the subsequent non-linear phase. 
We note that the time step is almost always set to the maximum value, $\Delta t = \Delta t_\mathrm{max}$, during the evolution of the short-timestep models. 
It only falls below this value, $\Delta t < \Delta t_\mathrm{max}$, near the end of the simulation, typically before it terminates due to numerical issues, as discussed below. 
Additionally, we perform simulations using doubled and halved values of the maximum time step for the models with $M_\mathrm{ini} = 14$ and $17\,\mathrm{M}_\odot$. 
A comparison of results with different time step values is presented in the Appendix \ref{sec:resolution_study}. 

For most models, we employ the same value for the parameter controlling the number of zones; \texttt{delta\_mesh\_coefficient=0.5}. 
We also try a halved value \texttt{delta\_mesh\_coefficient=0.25} for $M_\mathrm{ini}=14\,\mathrm{M}_\odot$ and $17\,\mathrm{M}_\odot$ model series, which roughly doubles the number of zones. 
The results are presented in the Appendix \ref{sec:resolution_study}. 
We also note that the influence of the mesh resolution on the pulsation properties of $1$--$5\,\mathrm{M}_\odot$ stars is recently examined by \cite{2025arXiv250113207L} in a more systematic manner. 
Ultimately, resolution studies as thorough as theirs would also be required for massive stars. 

\subsubsection{Final state and stability}\label{sec:final_states}
Massive and luminous RSGs are known to be unstable to radial pulsations. 
Accordingly, the short-timestep models generally exhibit growing radial pulsations, owing to time step values short enough to resolve their typical pulsation periods. 
As we shall discuss in more detail in Section \ref{sec:short-timestep}, models with lower initial masses tend to show regular pulsation cycles in the non-linear stage, whereas more massive models develop pulsations with much larger velocity amplitudes, eventually leading to the termination of the simulations. 
We refer to the former behaviour as a ``limit cycle'' and the latter as a ``pulsation runaway.''

For each model, we assess whether the simulation successfully completes the intended $100$ years of evolution by monitoring the ratio of surface velocity, $v_\mathrm{surf}$, to the escape velocity from the stellar surface, $v_\mathrm{esc}$. 
Models in which $v_\mathrm{surf} / v_\mathrm{esc} > 1$ at any point during the run are classified as “pulsation runaway” cases, while those maintaining $v_\mathrm{surf} / v_\mathrm{esc} < 1$ throughout and reach $t - t_\mathrm{ini} = 100\,\mathrm{yr}$ are considered “limit cycle” models. 
The final outcomes of all models are summarized in the rightmost columns of Tables \ref{table:model_description2} and \ref{table:model_description3}. 

\section{Results}
\subsection{Base models}
\subsubsection{Evolution}
\begin{figure}
\begin{center}
\includegraphics[scale=0.7]{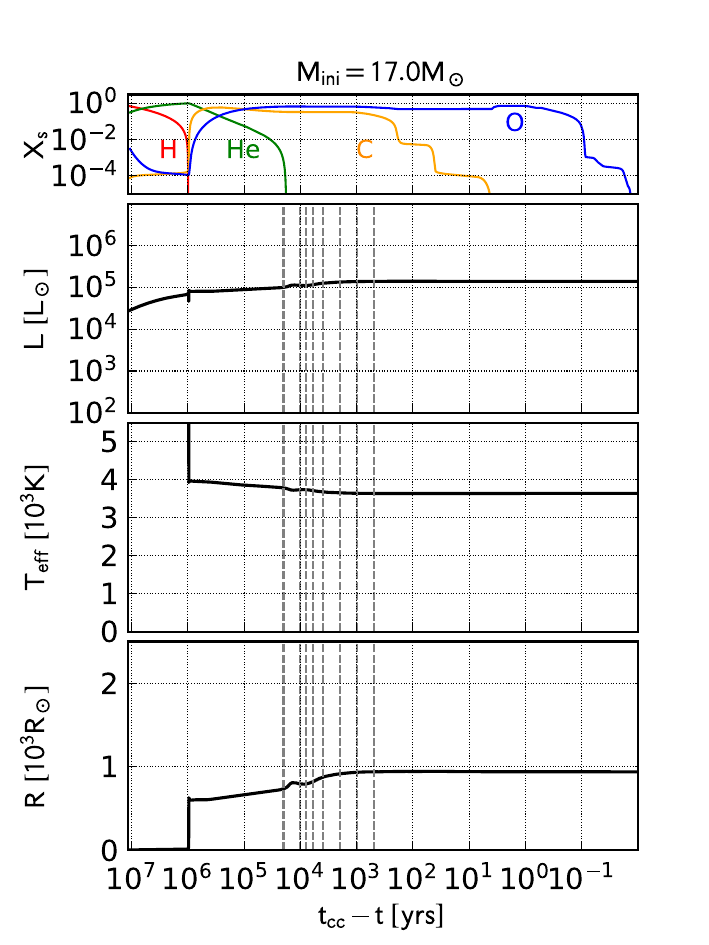}
\caption{Evolution of the properties of the base model with $M_\mathrm{ini}=17.0\,\mathrm{M}_\odot$. 
The central abundance (H, He, C, and O), the luminosity, the effective temperature, and the radius are plotted as a function of time prior to the core-collapse. 
The eight epochs of detailed simulations are designated by vertical dashed lines. 
}
\label{fig:evolution_M0170}
\end{center}
\end{figure}

The base models follow the standard evolutionary path of massive stars evolving into RSGs. 
Figure \ref{fig:evolution_M0170} shows the evolution of the $17\,\mathrm{M}_\odot$ model.
Following core H depletion at $t_\mathrm{cc}-t\simeq 10^6$ yr, the star transitions into the RSG phase: its radius expands to several hundred solar radii, and its surface temperature drops to around $3700\,\mathrm{K}$.
The radius continues to increase slightly during core He and C burning, eventually reaching $R \simeq 10^3\,\mathrm{R}_\odot$ by the end of core C burning.
These gradual changes in stellar structure are reflected in the properties of the expected radial pulsations, such as the pulsation period.
The figure also marks the eight epochs corresponding to the short-timestep models, which are used to further investigate the pulsation behaviour.
These epochs span from the end of core He burning to the end of core C burning.
After the final epoch, corresponding to $t_\mathrm{cc} - t \simeq 500$ yr, the global properties of the stellar envelope (e.g., radius and luminosity) remain nearly constant.
As a result, the pulsation characteristics of the envelope are expected to remain effectively unchanged beyond core C depletion.

As summarized in Table \ref{table:model_description}, the stellar radius at the time of iron core collapse generally increases with the initial mass $M_\mathrm{ini}$, ranging from $R = 710\,\mathrm{R}_\odot$ to $1032\,\mathrm{R}_\odot$. 
In contrast, the pre-supernova mass exhibits a more complex dependence on initial mass due to the intricacies of mass-loss processes throughout stellar evolution.
The core mass and pre-supernova luminosity, however, vary more monotonically with initial mass. 
The base models span a luminosity range of $\log_{10}(L/\mathrm{L}_\odot) = 4.9$–$5.2$, which encompasses the proposed upper luminosity limit for SNe-IIP progenitors. 
In the subsequent sections, we show that the seemingly constant luminosity observed during core C burning is, in fact, an artifact of the long time steps used in standard evolutionary calculations. 
Short-timestep models reveal physically meaningful luminosity variations during this phase.

\subsubsection{Linear perturbation analysis}
We first perform linear perturbation analysis for the base models after their main sequence stages. 
We use the open-source stellar oscillation code \textsc{Gyre} \citep{2013MNRAS.435.3406T,2018MNRAS.475..879T} for the linear analysis.  
We seek the non-adiabatic oscillation mode with the longest period by using a Magnus multiple shooting scheme (\texttt{MAGNUS\_GL2}) for stellar structures in each model. 
The linear analysis by \textsc{Gyre} assumes that the perturbations evolve with $\propto \exp(-i\omega \bar{t})$, where $\bar{t}$ is the time normalized by the global dynamical time;
\begin{equation}
    \bar{t}=t\qty(\frac{GM}{R^3})^{1/2},
\end{equation}
with $M$ and $R$ being the mass and radius of the star, and returns the real and imaginary parts of the non-dimensional $\omega$ for each oscillation mode. 
We focus on the fundamental radial oscillation mode (i.e., the degree is zero, $l=0$) with the period $P_\mathrm{gyre}$ and the growth rate $\eta_\mathrm{lin}$ given by
\begin{equation}
    P_\mathrm{gyre}=\frac{2\pi}{\mathrm{Re}[\omega]}\qty(\frac{GM}{R^3})^{-1/2},
\end{equation}
and 
\begin{equation}
    \eta_\mathrm{lin}=\frac{\mathrm{Im}[\omega]}{\mathrm{Re}[\omega]}.
\end{equation}

In general, the massive RSGs investigated in this study are known to be unstable to fundamental-mode radial pulsations. 
Consequently, the non-adiabatic linear perturbation analysis performed with {\sc Gyre} yields positive growth rates. Our short-timestep models indeed exhibit growing pulsations with periods consistent with those predicted by the linear perturbation analysis, as discussed below. 
However, there is a discrepancy between the growth rates of unstable radial pulsation modes obtained from the {\sc Mesa} simulations and those from the non-adiabatic linear analysis by {\sc Gyre}. 
This discrepancy is likely due to different approximations implemented in the linear analysis and numerical code. 
For instance, the non-adiabatic oscillation equations presented by \citet{2018MNRAS.475..879T} neglect perturbations to convective heating and cooling. 
Although {\sc Gyre} offers several treatments of convection, we opted to use the linear perturbation analysis primarily to identify the fundamental radial pulsation mode and determine its period and stability.

\subsubsection{Period-luminosity relation}
\begin{figure*}
\begin{center}
\includegraphics[scale=0.8]{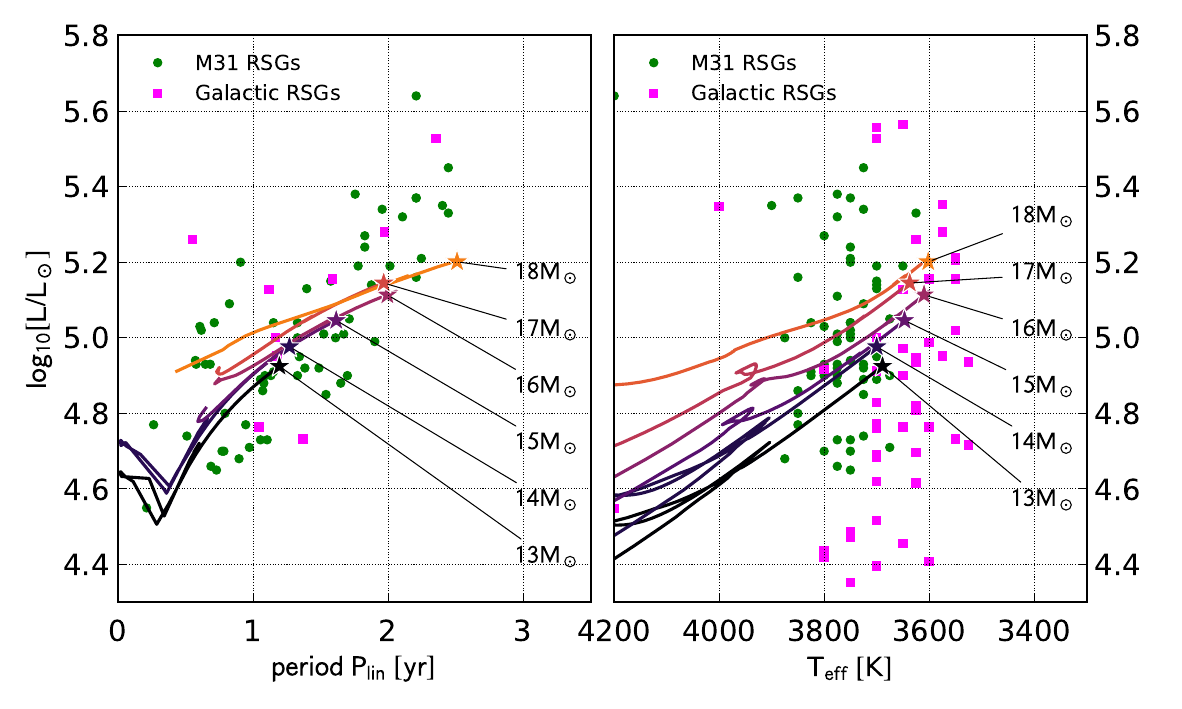}
\caption{Evolution of six base models with $M_\mathrm{ini}/\mathrm{M}_\odot=13$, $14$, $15$, $16$, $17$, and $18$ in the luminosity vs period (left) and luminosity vs surface temperature (right) plots. 
The evolutionary tracks beyond RSG stage are plotted as solid lines in the two panels. 
The end of the tracks (the iron core-collapse) are designated by star symbol. 
The RSGs in our Galaxy (circle) and the nearby galaxy M31 (square) are also plotted for comparison. 
}
\label{fig:obs_comparison}
\end{center}
\end{figure*}

We compare the properties of the base models with observations to assess the validity of key parameters. 
In Figure~\ref{fig:obs_comparison}, we show the evolutionary tracks of the 6 base models in the $L$--$P$ and $L$--$T_\mathrm{eff}$ planes. 
For comparison, we also include observational data for RSGs in the Milky Way and M31. 
The data are taken from \citet{2005ApJ...628..973L} for Galactic RSGs, \citet{2006MNRAS.372.1721K} for Galactic pulsating RSGs, and \citet{2016ApJ...826..224M} for pulsating RSGs in M31. 
The evolutionary tracks show overall agreement with the observed properties in both planes, supporting the validity of the chosen parameters, such as the mixing-length parameter, used in our models. 

\subsection{Short-timestep models}\label{sec:short-timestep}
In the following, we describe results of short-timestep model series. 

\subsubsection{Growth and saturation of radial pulsation}
\begin{figure}
\begin{center}
\includegraphics[scale=0.56]{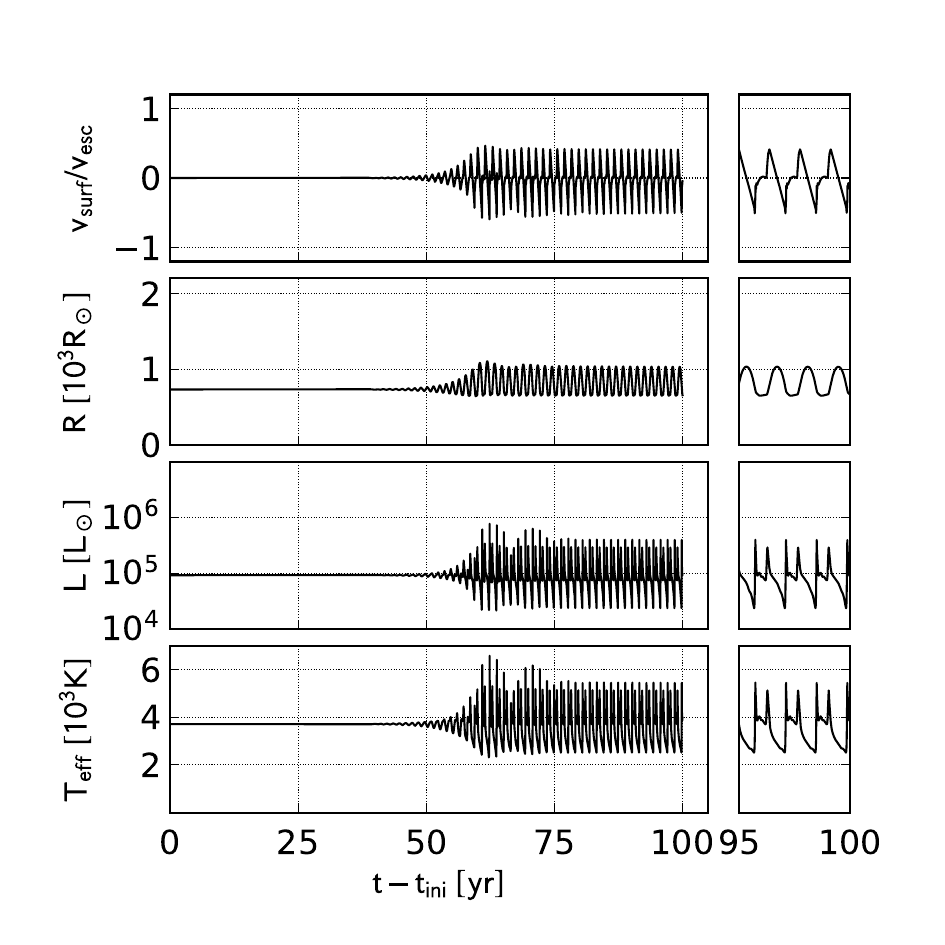}
\caption{Evolution of radial pulsation in \texttt{M0140\_model7}. 
{\it Left}: the surface velocity divided by the escape velocity, the radius, the luminosity, and the surface temperature are plotted as a function of time from top to bottom. 
{\it Right}: the up-close view of the profile in the last 5 years to clarify the pulse shape. 
}
\label{fig:pulsation_M0140_model7}
\end{center}
\end{figure}

Figure~\ref{fig:pulsation_M0140_model7} shows the evolution of model \texttt{M0140\_model7}, which represents a $14.0\,\mathrm{M}_\odot$ star approximately $10^3$ years before core collapse (see Table~\ref{table:model_description2}). 
As clearly seen in the figure, the radial perturbation, which is initially triggered by numerical noise, grows rapidly. 
During the growth phase ($t - t_\mathrm{ini} \lesssim 60$ yr), the amplitude of surface velocity oscillations increases exponentially, eventually reaching a significant fraction (about $30$--$40\%$) of the escape velocity. 
In this model, the initial escape velocity is $v_\mathrm{esc} = 82\,\mathrm{km\,s^{-1}}$. 
Simultaneously, the stellar radius, luminosity, and surface temperature oscillate around their initial values.

Such radial pulsations are suppressed in the base models due to the use of long time steps. 
In contrast, the short-timestep models successfully capture the unstable growth of radial pulsations predicted by the linear perturbation analysis. 
In this model, the analysis performed with {\sc Gyre} yields a fundamental pulsation period of $P_\mathrm{gyre} = 1.20$ yr. 
The surface properties shown in Figure~\ref{fig:pulsation_M0140_model7} exhibit pulsations at nearly the same period, confirming the consistency between the simulation and the linear analysis.

\begin{figure*}
\begin{center}
\includegraphics[scale=0.60]{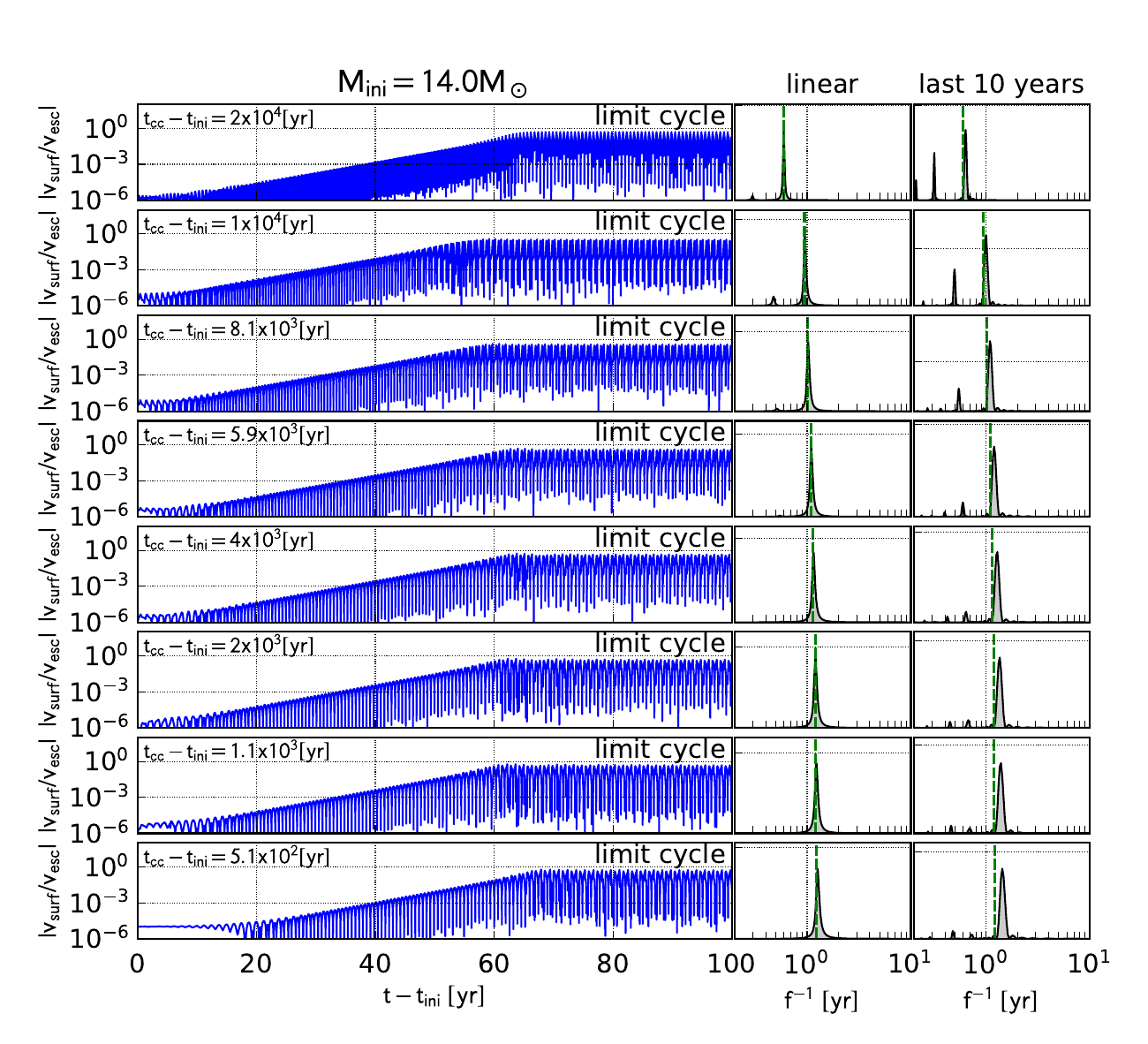}
\caption{Pulsation properties of \texttt{M0140} model series. 
The eight short-timestep models with the initial epochs ranging from $t_\mathrm{cc}-t_\mathrm{ini}=2\times 10^4$ to $5.1\times 10^2$ yr (see, Table \ref{table:model_description2}) are plotted from top to bottom. 
The left column presents the temporal evolution of the absolute values of the surface velocity divided by the escape velocity, $|v_\mathrm{surf}/v_\mathrm{esc}|$. 
The two columns on the right presents the results of the period analysis for the surface velocity evolution on the left. 
The periodograms for the linear phase and the last 10 years are presented. The vertical dashed lines indicate the periods from the linear analyses. 
}
\label{fig:pulsation_M140}
\end{center}
\end{figure*}

Figure \ref{fig:pulsation_M140} shows the pulsation properties of short-timestep models with $M_\mathrm{ini}=14.0\,\mathrm{M}_\odot$. 
All the models exhibit similar surface velocity evolutions to \texttt{M0140\_model7}; an exponential increase of the amplitude in the linear stage followed by oscillations with a finite amplitude in the non-linear stage. 
For a more quantitative comparison, we perform period analysis of this surface velocity evolution. 
We take time intervals during which the absolute value of the surface velocity is $0.01$--$10\%$ of the escape velocity, $10^{-4}<|v_\mathrm{surf}/v_\mathrm{esc}|<10^{-1}$, to study the oscillation modes in the linear regime. 
We also examine the non-linear pulsation period by taking the velocity evolution in the last 10 years. 
We use the standard Lomb-Scargle periodogram \citep{1976Ap&SS..39..447L,1982ApJ...263..835S} in \texttt{astropy} package (version 6.1.0; \citealt{2013A&A...558A..33A,2018AJ....156..123A,2022ApJ...935..167A}) to identify the period of the dominant oscillation mode in the frequency range of $f\in[f_\mathrm{min},f_\mathrm{max}]$ with $f_\mathrm{min}=0.1$ and $f_\mathrm{max}=10$ yr$^{-1}$. 
The obtained periods in the linear stage and the last 10 years are denoted by $P_\mathrm{lin}$ and $P_\mathrm{10yr}$.  
As is expected from the quite regular pulse profiles in Figure \ref{fig:pulsation_M0140_model7}, the period analysis unambiguously identifies the period of the dominant oscillation mode (see, Figure \ref{fig:pulsation_M140} below). 
The measured periods for all the short-timestep models are reported in Tables \ref{table:model_description2} and \ref{table:model_description3}.

The periodograms for the linear phase in Figure \ref{fig:pulsation_M140} are characterised by sharp peaks at frequencies consistent with the linear perturbation analysis by \textsc{Gyre}, which confirms that the simulations successfully capture the excitation and growth of the fundamental radial pulsation mode. 
All the models in this model series show quite regular pulses in the non-linear stage, in a similar way to Figure \ref{fig:pulsation_M0140_model7}.

The periodograms corresponding to the last 10 years of the velocity evolution are characterized by sharp peaks similar to those in the linear stage. 
The corresponding periods are, however, slightly longer than the values in the linear stage and the prediction by the linear perturbation analysis. 
Also, several periodograms exhibit small peaks at the periods shorter than those of the main peaks by a factor of $2$. 
These are caused by the non-linear pulse profiles. 
This shows that the power of the 1st overtone is generally smaller than that of the fundamental mode. 
Therefore, the growing pulsation energy of the fundamental mode should be dissipated and transported away rather than being transferred to other modes through non-linear couplings, although we cannot rule out non-linear coupling to non-radial modes. 


\subsubsection{Limit cycle\label{sec:limit_cycle}}
\begin{figure*}
\begin{center}
\includegraphics[scale=0.68]{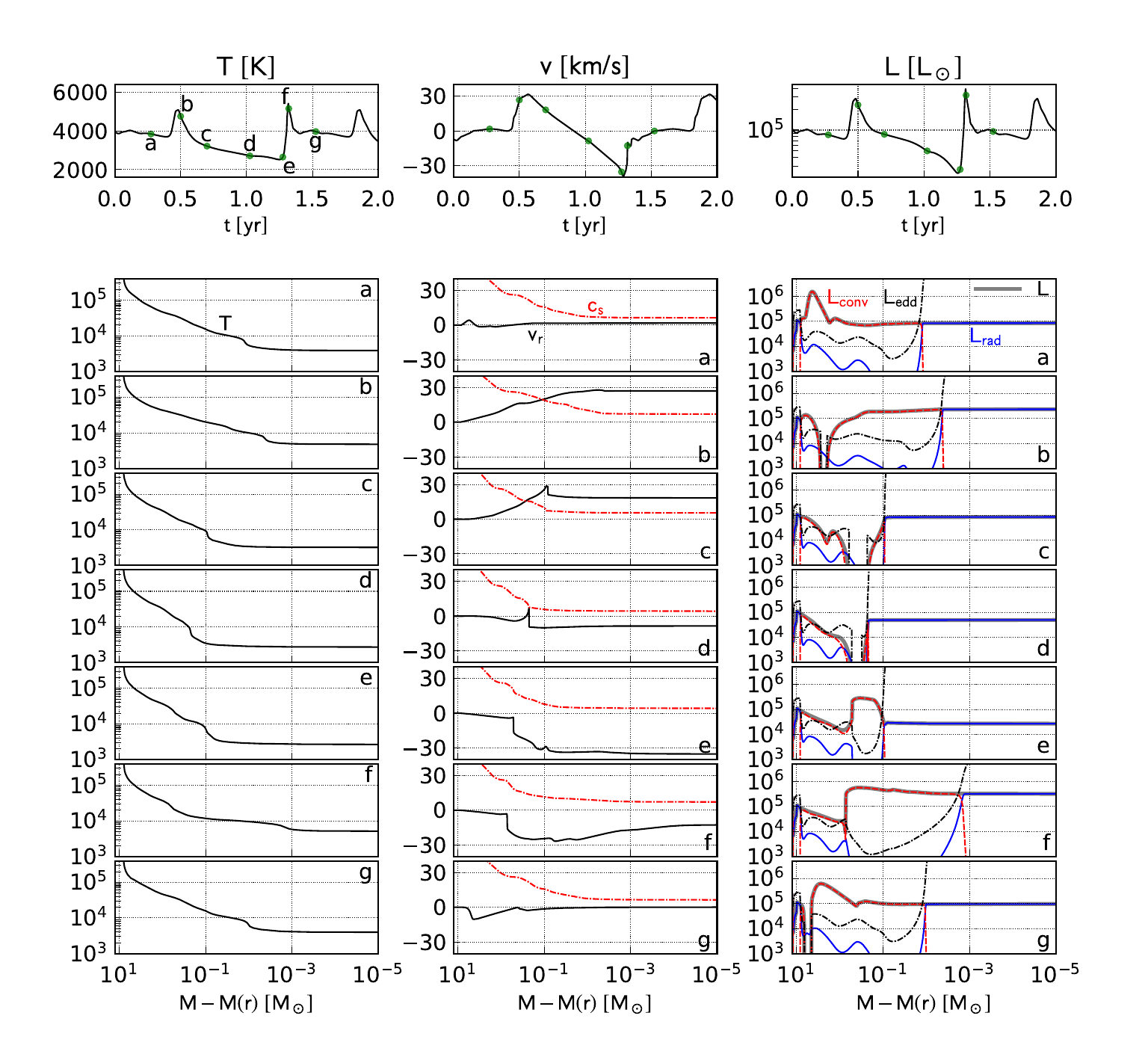}
\caption{Envelope structure during a non-linear cycle for \texttt{M0140\_model7}. 
The evolution and the distribution of the temperature (left), the velocity (centre), and the luminosity (right) are presented. 
In each column, the top panel presents the temporal evolution of the surface value in a single cycle of the pulsation. 
The remaining panels, (a) to (g), present the distributions of each quantity as a function of the mass coordinate from the surface, $M-M(r)$. 
The epochs corresponding to the panels (a) to (g) are designated in the temporal evolutions in the top row. 
In the radial velocity structures (centre), we also plot the distribution of the local sound speed, $c_\mathrm{s}$. 
In the luminosity structure (right), we plot the convective ($L_\mathrm{conv}$; dashed line), radiative ($L_\mathrm{rad}$; thin solid line), and total (thick solid line) luminosity distributions. 
We also plot the local Eddington luminosity $L_\mathrm{edd}$ (dash-dotted line). 
}
\label{fig:limit_cycle}
\end{center}
\end{figure*}
Here, we examine the envelope pulsation during the non-linear stage in greater detail.
As shown in Figure~\ref{fig:pulsation_M0140_model7}, model \texttt{M0140\_model7} settles into regular pulsation cycles shortly after entering the non-linear regime at $t - t_\mathrm{ini} \simeq 60\,\mathrm{yr}$.
The pulse profiles displayed in the right panels of Figure~\ref{fig:pulsation_M0140_model7} exhibit complex structures, featuring multiple peaks and slope changes within a single period, rather than smooth, sinusoidal variations.

This non-linear behaviour arises from a complex interplay between the amplification and dissipation of pulsation energy.
Even in the non-linear regime, the star emits radiation at a rate that, when averaged over long timescales, closely matches the core luminosity.
Thus, it is crucial to clarify when the pulsation energy is stored as kinetic energy in the envelope, and how this energy is eventually converted into radiation that escapes through the photosphere.

Figure~\ref{fig:limit_cycle} illustrates the physical quantities in the stellar envelope during non-linear pulsation.
We present the profiles as a function of the mass coordinate from the surface, $M - M(r)$, where $M(r)$ is the mass coordinate measured from the center. These profiles correspond to seven characteristic phases (labeled {\it a} to {\it g}) within a single pulsation cycle.
In phase ({\it a}), regions with positive radial velocities ($v_r > 0$) are located at $M - M(r) \simeq$ a few $\mathrm{M}_\odot$, significantly below the surface.
Consequently, the surface velocity (middle column panel) remains close to zero, $v_\mathrm{surf} \simeq 0$.
The temperature and density structures at this phase are therefore similar to those in the unperturbed state.

The temperature distribution (left column) monotonically increases with depth (i.e., increasing $M - M(r)$).
A relatively steep temperature gradient appears around $M - M(r) \simeq 10^{-2}\,\mathrm{M}_\odot$, above which the temperature drops from $T \simeq 8000\,\mathrm{K}$ to $T \simeq 5000\,\mathrm{K}$.
This temperature range crosses the ionization/recombination threshold of hydrogen, typically $6000$--$7000\,\mathrm{K}$.
Below this transition layer, free electrons contribute significantly to opacity, whereas above the layer, opacity drops down to significantly lower values.
This boundary therefore corresponds to the ``ionization front'' for neutral hydrogen. 

In this model, the ionization front consistently coincides with the interface between the inner convective zone and the outer radiative layers. 
The luminosity profiles (right column of Figure~\ref{fig:limit_cycle}) indicate that energy is transported by convection in the inner layers at a rate exceeding the local Eddington luminosity, $L_\mathrm{edd}$. 
This occurs because the high opacity in the ionized region inhibits efficient radiative diffusion, making convection the dominant mode of energy transport. 
In contrast, above the ionization front, the significantly lower opacity enables efficient energy transport via radiative diffusion.
This transition is reflected in the steep increase of the local Eddington luminosity from the ionization front toward the surface. 

The region with positive radial velocities in phase ({\it a}) corresponds to an outward-propagating acoustic wave.
As it travels into the lower-density outer layers, it accelerates the material, transporting energy outward and heating these layers.
This outward energy transfer drives the ionization front further from the centre and results in the emission of radiation from the stellar surface, producing the luminosity and temperature peaks observed in phase ({\it b}). 
Once the wave passes through the outer envelope, the region with $M - M(r) > 0.1\,\mathrm{M}_\odot$ begins expanding at supersonic speeds ($v_r > c_\mathrm{s}$). 
During this expansion, both adiabatic and radiative cooling reduce the temperature, causing the ionization front to recede inward (phase {\it c}). 
As a result, radiation pressure becomes ineffective at sustaining the outward acceleration. 
Since the expansion velocities remain below the local escape velocity, gravity eventually overcomes the motion and the envelope begins to contract. 
In phase ({\it d}), the radial velocities turn negative, indicating a supersonic infall. 
The infall is halted around a layer where the local sound speed becomes comparable to the infalling velocity. 
This is the location where the kinetic energy of the infalling material is dissipated. 
In phase ({\it e}), a convective region with high luminosity develops beneath $M - M(r) \simeq 0.1\,\mathrm{M}_\odot$, redistributing the dissipated energy to the overlying, still-infalling layers. 
This energy input heats the material and pushes the ionization front outward once again (phase {\it f}), producing a secondary set of luminosity and temperature peaks. 
As radiation penetrates and emerges through the infalling layers, it decelerates them and eventually halts further infall. 
After the radiation breakout, the residual kinetic energy of the infalling material is transmitted inward as an acoustic wave (phase {\it g}). 
This wave reflects at the core-envelope interface, which acts like a rigid boundary, producing a phase shift of $\pi$. 
The reflected wave initiates the next pulsation cycle, bringing the system back to phase ({\it a}). 

In summary, the pulsation energy is released in the form of radiation through two distinct mechanisms. 
The first is associated with the emergence of an outward-propagating, accelerating wave at the stellar surface, resulting in luminosity and temperature peaks accompanied by a positive surface velocity. 
The second occurs during the deceleration of infalling material, where the dissipation of kinetic energy also produces peaks in luminosity and temperature, this time associated with a negative surface velocity. 
In this particular simulation, the growth and dissipation of pulsation energy are well balanced, leading to regular pulsation cycles for several decades. 

\subsubsection{Pulsation runaway}
\begin{figure}
\begin{center}
\includegraphics[scale=0.56]{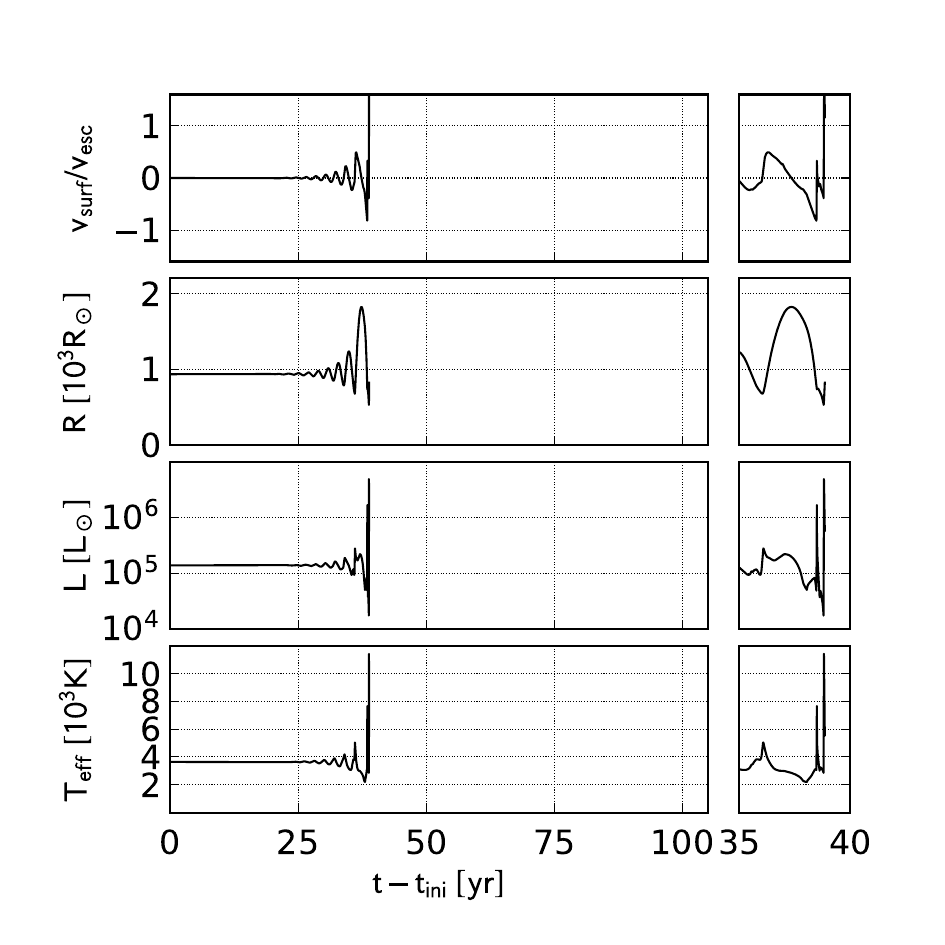}
\caption{Same as Figure \ref{fig:pulsation_M0140_model7}, but for \texttt{M0170\_model7}. 
}
\label{fig:pulsation_M0170_model7}
\end{center}
\end{figure}

The non-linear behaviour of the radial pulsation is remarkably different for higher mass models. 
The evolution of the surface properties for model \texttt{M0170\_model7} is presented in Figure \ref{fig:pulsation_M0170_model7}. 
While the oscillations grow from small perturbations as in the case of model \texttt{M0140\_model7}, 
the oscillation amplitudes eventually become so large at $t-t_\mathrm{ini}\simeq 35\,\mathrm{yr}$ that the surface velocity exceeds the escape velocity. 
This makes the time step $\Delta t$ of the computation even shorter and eventually leads to the termination of the run. 
Shortly before the termination, the stellar radius reaches a minimum. 
The luminosity and temperature instead show a spike with their peak values of $L\simeq 4\times 10^6\,\mathrm{L}_\odot$ and $T\simeq 10^4\,\mathrm{K}$. 
Since the stellar surface is expanding at velocities faster than the escape velocity at the termination of the run, it is likely associated with some kinds of mass ejection. 

\begin{figure*}
\begin{center}
\includegraphics[scale=0.60]{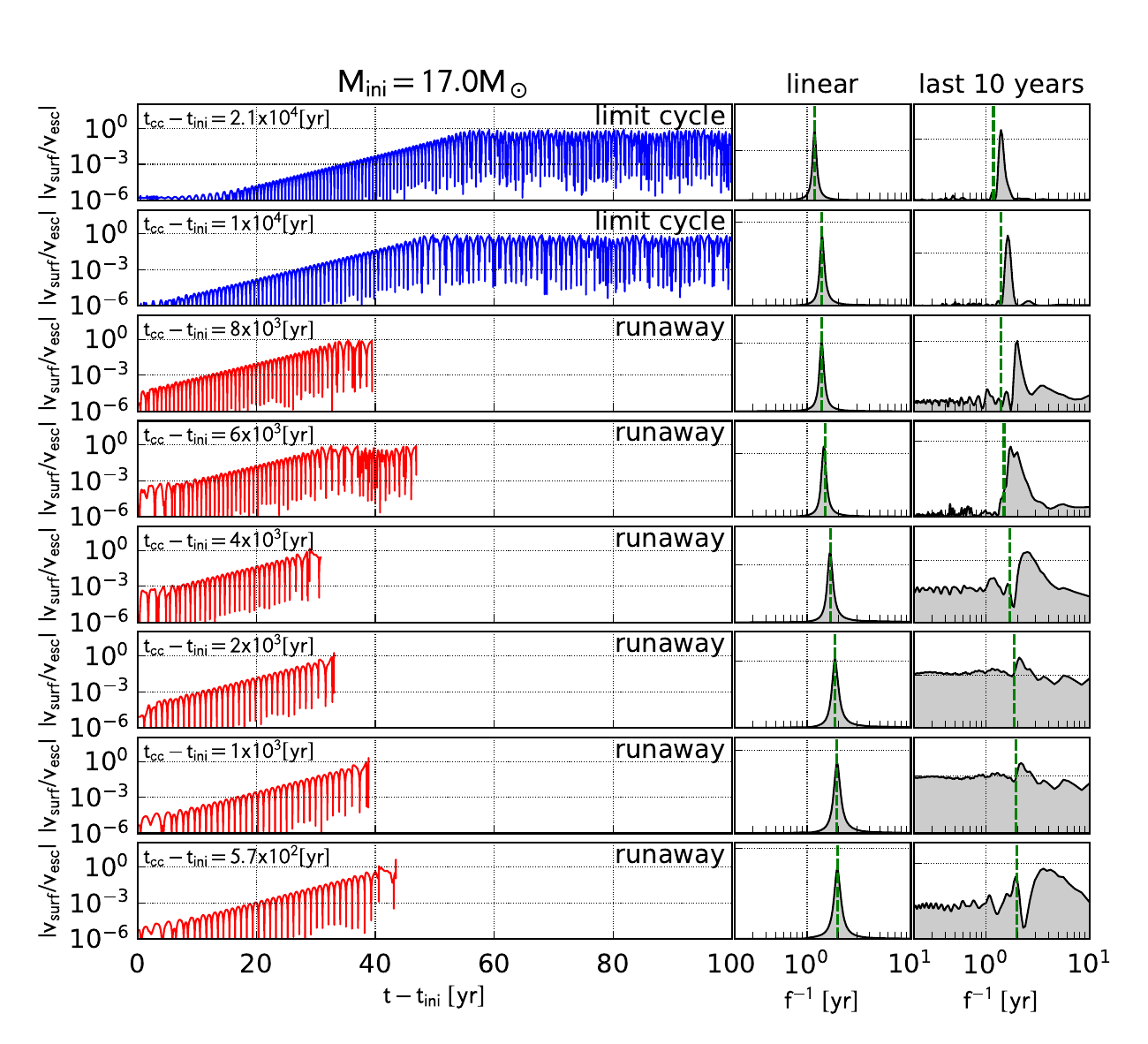}
\caption{Same as Figure \ref{fig:pulsation_M140}, but for \texttt{M0170} model series. 
}
\label{fig:pulsation_M170}
\end{center}
\end{figure*}

Figure \ref{fig:pulsation_M170} presents the surface velocity evolution for the short-timestep models with $M_\mathrm{ini}=17\,\mathrm{M}_\odot$. 
The first two models (\texttt{M0170\_model1} and \texttt{M0170\_model2}) complete the evolution for 100 yr. 
These models, however, exhibit irregular surface velocity evolution in their non-linear stages. 
This is in contrast to the less massive models (Figure \ref{fig:pulsation_M140}) showing complex, but fairly regular pulsation cycles. 
Therefore, although we classify these two models as limit cycle cases according to our criterion (Sec. \ref{sec:final_states}), they may experience pulsation runaway at some point on the course of further evolution. 
The remaining six models, which start the short-timestep calculations at later epochs, 
including the model shown in Figure \ref{fig:pulsation_M0170_model7}, clearly experience the pulsation runaway. 
The last four models with $t_\mathrm{ini}-t_\mathrm{cc}<4\times 10^3\,\mathrm{yr}$ are terminated around the end of the linear stage. 
For the two models with $t_\mathrm{cc}-t_\mathrm{ini}=8$ and $6\times 10^3\,\mathrm{yr}$, on the other hand, the surface velocity evolutions show irregular oscillations before exceeding the escape velocity. 

\begin{figure*}
\begin{center}
\includegraphics[scale=0.68]{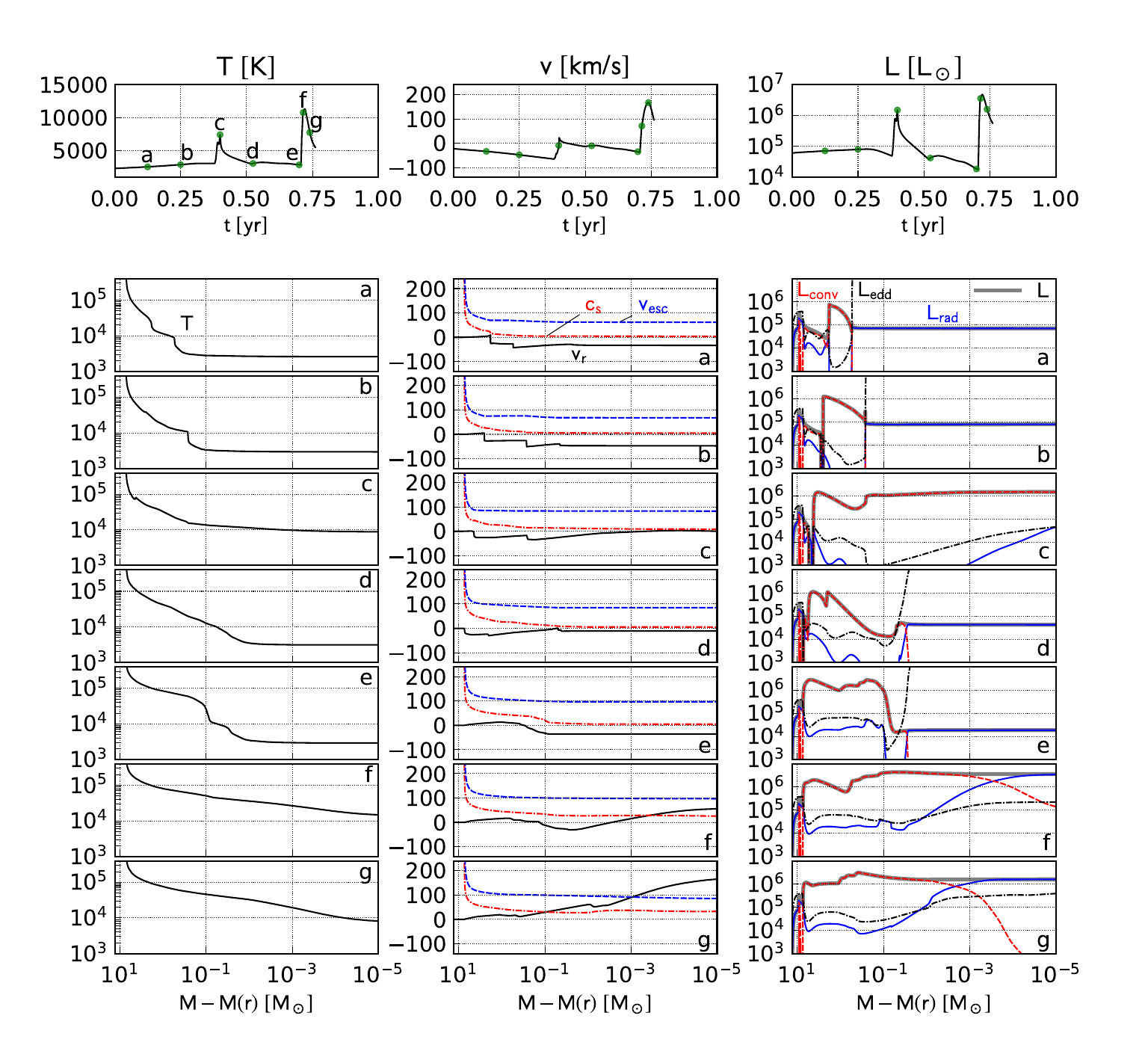}
\caption{Same as Figure \ref{fig:limit_cycle}, but for the pulsation runaway case. 
}
\label{fig:ejection}
\end{center}
\end{figure*}

In order to exactly understand how the pulsation runaway happens, we investigate the structure of the stellar envelope leading to the runaway in Figure \ref{fig:ejection}. 
We only focus on the time interval less than 1 year before the surface velocity exceeds the escape velocity. 
As seen in the top row of the figure, the temperature, the velocity, and the luminosity exhibit sudden increases twice. 
In a similar way to the limit cycle case in Section \ref{sec:limit_cycle}, we specify 7 characteristic phases in the temporal evolution of the envelope, including the luminosity peaks with $L>10^6\,\mathrm{L}_\odot$. 
In phase ({\it a}), the envelope has already been infalling with supersonic velocities after the expansion and cooling. 
The ionization front and the convective-radiative interface are at the same position in the deep interior of the envelope. 
In a similar way to the limit cycle case, the continued infall of the envelope builds up a region with high luminosity (phase {\it b}). 
This excessive energy is transported via convection, heating and ionizing materials ahead (phase {\it c}). 
This is followed by the release of the energy in the form of radiation, making a luminosity peak. 
After releasing the accumulated energy, the ionization front propagates back to deeper layers. 
However, unlike the limit cycle case in Figure \ref{fig:limit_cycle}, a high-luminosity layer remains at $M-M(r)\simeq$ several $\mathrm{M}_\odot$ with a continued infall (phase {\it d}). 
The radial velocity of this high-luminosity layer turns to positive and the ionization front again propagates toward the surface (phase {\it e}). 
In phase ({\it f}), the outwardly transported energy heats the envelope to temperatures higher than the hydrogen ionization temperature. 
The surface temperature at this phase is higher than $10^4\,\mathrm{K}$, i.e., the entire envelope is ionized. 
As a consequence, a radiative layer with the luminosity exceeding the local Eddington luminosity, $L_\mathrm{rad}>L_\mathrm{edd}$ appears. 
This super-Eddington radiative layer keeps accelerating outer layers above $M-M(r)>10^{-3}\,\mathrm{M}_\odot$, whose velocities eventually exceed the escape velocity (phase {\it g}).

\subsection{Pulsating RSGs and possible mass ejection}

\subsubsection{Limit cycle or runaway}
We briefly discuss the conditions for the pulsation runaway. 
As seen in the previous subsections, the non-linear behaviour of the pulsating RSG envelope is highly complicated due to the coupling between convection and pulsation. 
Therefore, the behaviour would be strongly dependent on the treatment of the convection in the evolutionary computation and relevant parameters. 
Nevertheless, it is worth clarifying how exactly the growing pulsation saturates in the present \textsc{Mesa} simulations. 

We consider the non-linear pulsations of the RSG envelope in the limit cycle and runaway cases in the following simplified manner. 
As we have seen in Section \ref{sec:short-timestep}, the kinetic energy of the pulsating layer appears to be dissipated at a specific layer and transported to the stellar surface. 
We thus consider that a part of the envelope above the layer oscillates with a characteristic velocity $v_\mathrm{osc}(>0)$. 
The remaining inner part of the star is assumed to be at rest. 
In the limit cycle case, the surface velocity exhibits a finite-amplitude oscillations and thus the characteristic velocity is simply given by the maximum velocity amplitude. 
In the runaway case, on the other hand, we define the velocity as the maximum infall velocity at the surface shortly before the the termination of the run. 
As we have seen in Figures \ref{fig:limit_cycle} and \ref{fig:ejection}, the characteristic layer separating the outer oscillating and inner non-oscillating parts appears around the region with the local sound velocity comparable to the characteristic velocity, $c_\mathrm{s}\sim v_\mathrm{osc}$. 
In the infalling phase, the materials above the critical radius has negative radial velocities of $-v_\mathrm{osc}$ and dissipate their kinetic energy at the critical radius. 
The position of the critical radius is thus determined by the balance between the thermal pressure $p$ at the interface of the inner and outer parts, and the ram pressure $\rho v_\mathrm{osc}^2$ of the infalling materials, where $\rho$ is the density around the layer. 
This balance, $p\sim \rho v_\mathrm{osc}^2$, leads to the condition $c_\mathrm{s}\sim v_\mathrm{osc}$, since the scaling relation $p/\rho\sim c_\mathrm{s}^2$ holds by neglecting numerical factors. 
Denoting the critical radius by $R_\mathrm{cr}$ and assuming that the infall stops at the radius, the energy dissipation rate for the infalling materials is given by 
\begin{equation}
L_\mathrm{diss}=2\pi R_\mathrm{cr}^2\rho v_\mathrm{osc}^3\sim 2\pi R_\mathrm{cr}^2pv_\mathrm{osc}.
\label{eq:L_diss}
\end{equation}

Alternatively, a similar luminosity can be obtained as follows. 
When the envelope is accelerated and expanding, we consider an acoustic wave traveling from the core-envelope boundary through the envelope. 
For a pressure deviation $\delta p$ from the background distribution, the acoustic power carried by the wave at radius $r$ is given by
\begin{equation}
    L_\mathrm{wave}=4\pi r^2\frac{\delta p^2}{\rho c_\mathrm{s}}.
\end{equation}
When the outwardly propagating wave accelerates the envelope to the characteristic velocity $v_\mathrm{osc}$, the non-linear effect becomes important at the critical radius, where $c_\mathrm{s}\sim v_\mathrm{osc}$. 
Because the wave speed becomes supersonic beyond the layer, the pressure deviation becomes comparable to the background value $\delta p\sim p$, at the critical radius $r=R_\mathrm{ch}$. 
Again using the relation $p\sim \rho c_\mathrm{s}^2$ neglecting some numerical factors, the acoustic power at the critical radius yields
\begin{equation}
    L_\mathrm{wave}=4\pi R_\mathrm{cr}^2\rho c_\mathrm{s}^3\simeq 4\pi R_\mathrm{cr}^2\rho v_\mathrm{osc}^3,
\end{equation}
which gives a similar value to the dissipated luminosity above, Eq. (\ref{eq:L_diss}). 

In model \texttt{M140\_model7} with the velocity amplitude $v_\mathrm{osc}=30\,\mathrm{km}\,\mathrm{s}^{-1}$, the critical radius and the density at the radius are found from the unperturbed structure; $R_\mathrm{cr}=5\times 10^2\,R_\odot$ and $\rho=2\times 10^{-8}\,\mathrm{g}\,\mathrm{cm}^{-3}$. 
The corresponding luminosity is calculated to be
\begin{equation}
    L_\mathrm{diss}\simeq 10^6\epsilon 
    \left(\frac{R_\mathrm{cr}}{500\,R_\odot}\right)^2
    \left(\frac{\rho_\mathrm{cr}}{2\times 10^{-8}\,\mathrm{g}\,\mathrm{cm}^{-3}}\right)
    \left(\frac{v_\mathrm{osc}}{30\,\mathrm{km}\,\mathrm{s}^{-1}}\right)^3\,\mathrm{L}_\odot,
\end{equation}
where $\epsilon$ is an efficiency factor absorbing the neglected numerical factors. 
In the right column of Figure \ref{fig:limit_cycle}, the peak surface luminosity reaches $L_\mathrm{surf}\simeq 4\times 10^5\,\mathrm{L}_\odot$ and the convective luminosity around the critical radius is even higher, $L_\mathrm{conv}\simeq 7\times 10^5\,\mathrm{L}_\odot$ at the epoch (phase {\it f}). 
Therefore, Eq. (\ref{eq:L_diss}) with the numerical efficiency factor of $\epsilon\simeq 0.4$--$0.7$ gives good estimates for the surface and inner convective luminosities. 
The efficiency factor close to unity suggests an efficient release of the dissipated energy and also reassures that the two-layer approximation well describes the dissipation process. 
The stellar radius of this model in the unperturbed state is $R=734\,R_\odot$. 
For releasing the dissipated luminosity $L_\mathrm{diss}$ estimated above through the original radius, the effective temperature should be
\begin{equation}
    T_\mathrm{eff}=\left(\frac{L_\mathrm{diss}}{4\pi \sigma_\mathrm{SB}R^2}\right)^{1/4}=5.7\times 10^3\,\mathrm{K},
    \label{eq:Teff}
\end{equation}
from the Stefan-Boltzmann law ($\sigma_\mathrm{SB}$ is the Stefan-Boltzmann constant), where we have assumed an efficiency factor of $\epsilon=0.5$ and thus $L_\mathrm{diss}=5\times 10^5\,\mathrm{L}_\odot$ from Equation (\ref{eq:L_diss}). 
This surface temperature is slightly lower than the Hydrogen ionization/recombination temperature. 
This estimate suggests that there is a thin layer dominated by neutral hydrogen at least around the surface of the star, i.e., the optical depth close to the outer boundary value of $\tau\sim 2/3$--$1$. 

In the runaway case of model \texttt{M0170\_model7}, the characteristic velocity is set to be the maximum infalling velocity shortly before the termination ($v_\mathrm{osc}=60\,\mathrm{km}\,\mathrm{s}^{-1}$; the top-middle panel of Figure \ref{fig:ejection}). 
The critical radius and the density are found to be $R_\mathrm{cr}\simeq 220\,R_\odot$ and $\rho=9\times 10^{-8}\,\mathrm{g}\,\mathrm{cm}^{-3}$. 
The luminosity arising from the dissipation of this infalling gas flow is estimated to be 
\begin{equation}
 L_\mathrm{diss}\simeq 7\times 10^6\epsilon 
    \left(\frac{R_\mathrm{cr}}{220\,R_\odot}\right)^2
    \left(\frac{\rho_\mathrm{cr}}{9\times 10^{-8}\,\mathrm{g}\,\mathrm{cm}^{-3}}\right)
    \left(\frac{v_\mathrm{osc}}{60\,\mathrm{km}\,\mathrm{s}^{-1}}\right)^3\,\mathrm{L}_\odot. 
\end{equation}
In Figure \ref{fig:ejection}, the maximum convective luminosity at the appearance of the super-Eddington layer is $L_\mathrm{conv}=4.4\times 10^6\,\mathrm{L}_\odot$, which again indicates an efficiency factor around $\epsilon\simeq 0.6$. 
A similar surface temperature estimate as in Equation (\ref{eq:Teff}) gives
\begin{equation}
    T_\mathrm{eff}\simeq 8.6\times 10^3\,\mathrm{K},
\end{equation}
for $R=935\,\mathrm{R}_\odot$ and $L_\mathrm{diss}=4.4\times 10^6\,\mathrm{L}_\odot$. 
In other words, this case produces ionized surface layer. 

The above considerations are based on a simplified two-layer model.
Nevertheless, this model quantitatively captures the key physical processes occurring in the oscillating envelope. 
When the pulsations grow to sufficiently large velocity amplitudes, the dissipation of pulsation energy leads to ionization of the entire envelope. 
As a result of this ``ionization breakout'', a super-Eddington radiative layer forms and is accelerated to velocities exceeding the surface escape velocity. 
In the present \textsc{Mesa} simulations, this process ultimately leads to the termination of the run. 

\subsubsection{Pulsation properties}\label{sec:condition}
As we have described above, the saturation of the growing radial pulsation in our short-timestep models can be realized in two different regimes; limit cycle and runaway cases. 
The results summarized in Tables \ref{table:model_description2} and \ref{table:model_description3} suggest that more massive RSGs with shorter lifetimes until the iron core-collapse are more likely to experience larger-amplitude radial pulsations accompanied by runaway. 
Our numerical experiments suggest an initial mass threshold around $M_\mathrm{ini}=16\,\mathrm{M}_\odot$, above which the pulsation runaway happens in the almost entire C burning stage. 

\begin{figure}
\begin{center}
\includegraphics[scale=0.7]{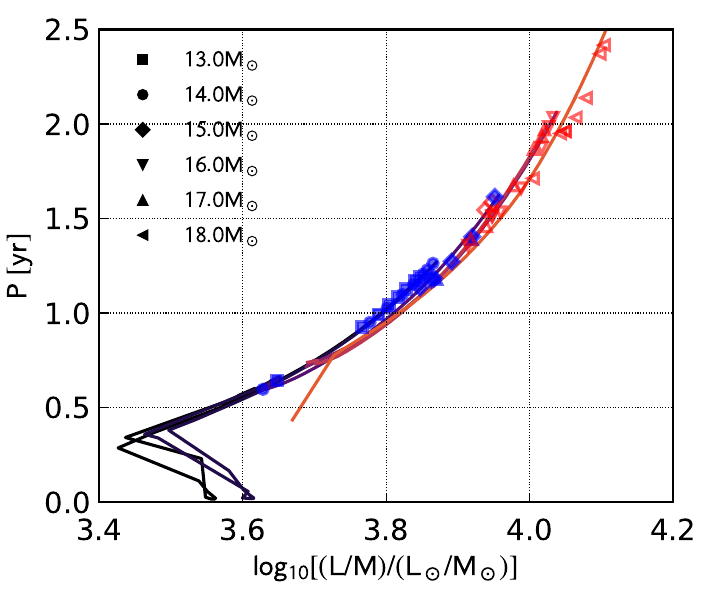}
\caption{Relation between the numerically obtained oscillation periods and the core luminosity-to-mass ratio for selected short-timestep models. 
The limit cycle and runaway cases are presented with filled (blue) and open (red) symbols, respectively. 
The $P$--$L/M$ relations predicted by the linear perturbation analysis are also plotted in the same manner as in Figure \ref{fig:obs_comparison}
}
\label{fig:period_LM}
\end{center}
\end{figure}

The pulsation properties of RSG envelopes are expected to be scaled with the luminosity-to-mass ratio, $L/M$. 
\cite{1965ApJ...142.1649G} showed that the oscillation period of a stellar atmosphere scales as $P\propto R^2/M$ by using a polytropic star model with an index $3/2$. 
\cite{1997A&A...327..224H} then examine the dependence of the pulsation periods of their RSG models on $L/M$, assuming almost constant effective temperature $L\propto R^2T_\mathrm{eff}^4$. 
The pulsation periods of \cite{1997A&A...327..224H} models certainly show a strong correlation with $L/M$. 
\cite{2010ApJ...717L..62Y} empirically found the growth rate $\eta$ of the amplitude of the radial pulsation better correlate with the following quantity,
\begin{equation}
    \eta\propto R^2 M^{-1}\tau_\mathrm{KH}^{-0.313},
\end{equation}
with $\tau_\mathrm{KH}$ being the Kelvin-Helmholtz time scale.

Motivated by these studies, we present the relation between the pulsation period $P$ and the luminosity-to-mass ratio $L/M$ using our models in Figure \ref{fig:period_LM}. 
The pulsation periods are determined by the period analysis and are summarised in Tables \ref{table:model_description2} and \ref{table:model_description3}.  
Along with the numerical results, we also show the $P$-$L/M$ relations from the linear perturbation analysis for the base models. 
Despite the different initial masses, the models exhibit a tight correlation in the $L/M$--$P$ plane up to $L/M\simeq 10^{4.1}\mathrm{L}_\odot/\mathrm{M}_\odot$. 
It is also noted that the numerically obtained $L/M$--$P$ relation is consistent with the prediction by the linear perturbation analysis. 

The limit cycle and runaway cases in different model series are well distinguished in this plane, although some models near the boundary are not clearly separated. 
We then empirically determine the following condition for pulsation runaway in terms of the luminosity-to-mass ratio,
\begin{equation}
    \log\qty(\frac{L/\mathrm{L}_\odot}{M/\mathrm{M}_\odot})\gtrsim 3.9.
    \label{eq:stability}
\end{equation}
The stellar mass at the late burning stages are in a narrow interval around $11$--$12\,\mathrm{M}_\odot$ for models close to this boundary, this condition can be expressed as a luminosity upper limit,
\begin{equation}
    \frac{L}{\mathrm{L}_\odot}\lesssim 10^{5}\qty(\frac{M}{12\,\mathrm{M}_\odot}),
\end{equation}
for avoiding pulsation runaway. 
Because of the tight $L/M$--$P$ relation, the condition can also be expressed as a condition for the pulsation period; $P\lesssim 1.7\,\mathrm{yr}$. 

\cite{1997A&A...327..224H} have shown that the growth rate of the radial oscillation steeply increases for $L/\mathrm{L}_\odot>10^{5.1}$ and thus the above luminosity upper limit derived with our larger model grid is similar to their results. 
We shall see that this luminosity is also similar to observational constraints on SNe-IIP progenitors in the next section. 
Since the stellar luminosity is a rapidly increasing function of the star mass, stars more massive than the initial mass range explored in this study would be more likely to lead to pulsation runaway, while less massive stars would be more stable. 

We note that the criterion on the luminosity-to-mass ratio depends on several parameters. 
For example, the stellar mass in the late evolutionary stage is influenced by the adopted mass-loss prescription. 
Although we have employed the default mass-loss prescription from the \texttt{20M\_pre\_ms\_to\_core\_collapse} inlists in the \textsc{Mesa} test suite (Section \ref{sec:mass_loss}), the treatment of stellar wind mass loss remains a topic of active debate \citep[e.g.,][]{2020MNRAS.492.5994B, 2023MNRAS.524.2460B,  2023A&A...676A..84Y,2024A&A...686A..88A,2024A&A...681A..17D}. 
Adopting alternative mass-loss prescriptions and/or varying the efficiency parameter can lead to different final stellar masses \citep[e.g.,][]{2017A&A...603A.118R,2024ApJ...964L..23R}, and hence different values of $L/M$. 
Even when the critical threshold in the luminosity-to-mass ratio $L/M$ remains unchanged, the initial mass range for limit cycle/pulsation runaway is consequently affected. 

In addition, the luminosity of a star in its advanced nuclear burning stages is primarily determined by its core mass, which develops after the core H depletion. 
The formation and subsequent growth of the core are strongly influenced by internal mixing processes, particularly near the core-envelope boundary. 
As a result, the initial stellar mass corresponding to the condition $L/M\simeq 10^{3.9}\mathrm{L_\odot}/\mathrm{M}_\odot$ also depends on parameters such as the mixing-length, convective overshooting, and other related prescriptions. 

\subsubsection{Onset of runaway phase and initial mass dependence}
\begin{figure}
\begin{center}
\includegraphics[scale=0.65]{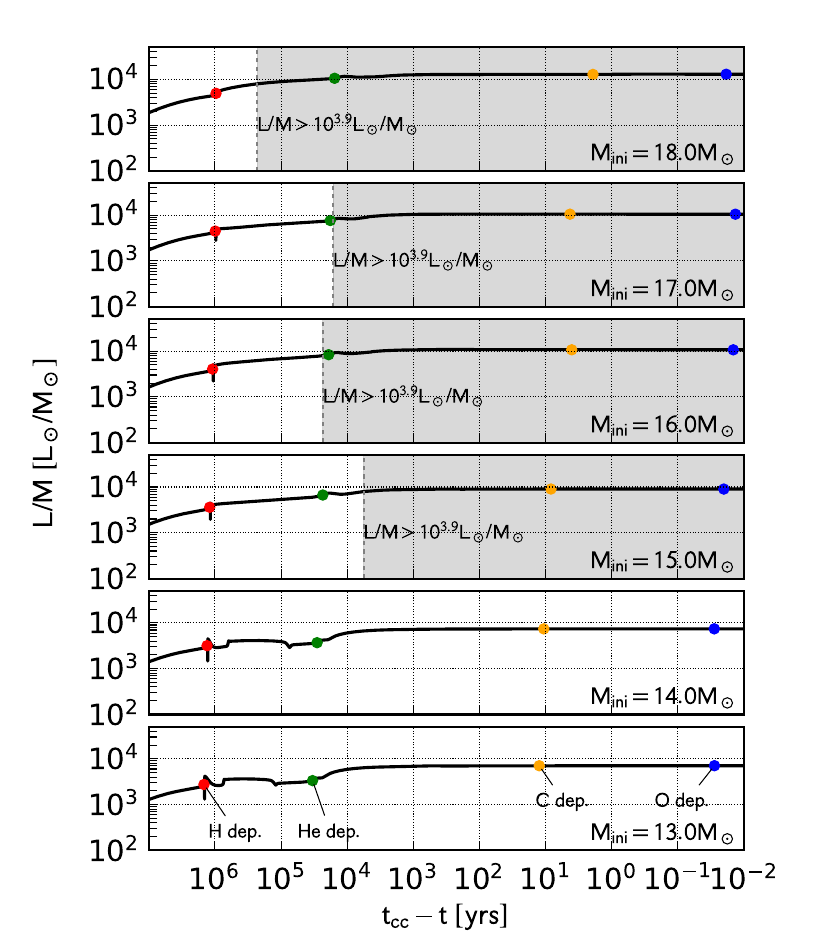}
\caption{Luminosity-to-mass ratio as a function of time until core-collapse for six base models with $M_\mathrm{ini}=13$--$18\,M_\odot$ (from bottom to top). 
In each panel, the circles represent the epochs of the hydrogen, helium, carbon, and oxygen depletion at the centre (from left to right). 
In shaded regions, the models satisfy the empirical condition $L/M>10^{3.9}\mathrm{L}_\odot/\mathrm{M}_\odot$ (Eq. \ref{eq:stability}) for pulsation runaway. 
}
\label{fig:evolution_L_over_M}
\end{center}
\end{figure}

We also consider the possibility of pulsation runaway occurring in earlier evolutionary stages. 
Although our short-timestep models primarily cover the C burning phase of the corresponding base models (Figure \ref{fig:evolution_M0170}), this does not necessarily preclude the onset of pulsation runaway at earlier epochs. 
The empirical condition for pulsation runaway, given by a luminosity-to-mass ratio of $L/M > 10^{3.9}\mathrm{L}_\odot/\mathrm{M}_\odot$, can serve as a useful guidance for identifying when a star is likely to enter the runaway phase. 
The temporal evolution of $L/M$ in the base models (Figure \ref{fig:evolution_L_over_M}) indicates that more massive stars satisfy the $L/M$ condition at earlier stages, while the less massive two models do not satisfy the condition during the entire evolution. 
This is expected, given the strong dependence of stellar luminosity on initial mass, $L \propto M^\alpha$ with $\alpha \sim 3$–$4$. 
In particular, the most massive model in our model grid, with $M_\mathrm{ini} = 18\,\mathrm{M}_\odot$, satisfies the $L/M$ condition already during the helium burning stage. 
While our model grid is limited to initial masses in the range $M_\mathrm{ini} = 13$–$18\,\mathrm{M}_\odot$, more massive stars ($M_\mathrm{ini}>18\,\mathrm{M}_\odot$) are likely to enter the runaway phase at even earlier stages. 
For instance, \cite{2020ApJ...902...63J} examined the radial pulsations of RSGs with $M_\mathrm{ini} = 18$–$23\,\mathrm{M}_\odot$ in the linear regime. 
Their \textsc{Mesa}-based hydrodynamic simulations start at core helium depletion (defined as a core helium mass fraction below $10^{-4}$) and demonstrate linear growth of the fundamental mode pulsation. 
Although non-linear pulsation behaviour was not investigated, these models could potentially exhibit pulsation runaway after saturation. 
However, the exact timing at which the $L/M$ condition is met depends on various factors, such as wind efficiency, as previously discussed. 
This highlights the need for further investigations into how parameters such as steady mass loss, convection, and initial mass influence the onset of pulsation runaway.

Very recently,\footnote{While this work was under peer review} \cite{2025arXiv250811077B} and \cite{2025arXiv250811088L} presented $10.5$, $12.5$, and $15\,\mathrm{M}_\odot$ RSG models using numerical setups similar to ours. 
Their study focuses on the pulsational behaviour of the model during the non-linear regime and its implications for the subsequent SN explosion. 
They investigated the growth and saturation of radial pulsations at the end of carbon burning and highlighted the importance of changes in the ionization structure over the pulsation cycle. 
At the time of core collapse, their $15\,M_\odot$ model has a mass and luminosity of $M=12.2\,\mathrm{M}_\odot$ and $L = 1.15 \times 10^5\,\mathrm{L}_\odot$, respectively, yielding a luminosity-to-mass ratio of $\log(L/M/(\mathrm{L}_\odot/\mathrm{M}_\odot)) = 3.97$ \citep{2025arXiv250811077B}. 
According to our empirical criterion, this value would place the model in the pulsation-runaway regime. 
On the other hand, the finite-amplitude pulsation in their models appears similar to our limit cycle cases. 
This discrepancy may simply indicate that the threshold value of $L/M$ does not perfectly distinguish the two cases, as in our model grids (Figure \ref{fig:period_LM}). 
Instead, it may arise from differences in the adopted parameters for convection and other physical treatments. For instance, their model assumes a lower mixing-length parameter, $\alpha_\mathrm{mlt} = 1.5$, and employs the \texttt{MLT++} option, which enhances convective energy transport. 
This underscores the need for further investigation into how such parameters affect the pulsational properties of RSGs.

How RSGs with different initial masses enter the pulsation runaway phase is particularly important, as the associated eruptive mass loss shedding a considerable fraction of the H-rich envelope may significantly shorten their lifetimes in the RSG phase. 
This effect could be reflected in the high-luminosity end of the RSG luminosity function. 
\cite{2023ApJ...942...69M} presented the luminosity functions of RSGs in M31 and M33, finding that the fraction of the most luminous RSGs is consistent with the evolutionary models of \cite{2012A&A...537A.146E}, who adopted enhanced mass-loss rates for stars approaching the Eddington limit. 
Further exploration of RSG pulsation properties across a wider parameter space, combined with robust comparisons to observed luminosity functions, may be critical for calibrating the physical parameters that govern the onset of pulsation runaway.

\subsubsection{Pulsation-driven mass-loss}
In the following, we discuss the possibility of the eruptive mass-loss repeatedly driven by the pulsation runaway, as is observed in massive RSG models above. 
One of the important quantities is how much mass is ejected in a single episode of the pulsation runaway. 
The $17\,\mathrm{M}_\odot$ model in Figure \ref{fig:ejection} shows that the outer envelope with the mass of $\sim 10^{-3}\,\mathrm{M}_\odot$ is accelerated to radial velocities faster than the escape velocity. 
A natural consequence of the envelope acceleration beyond the escape velocity would be mass ejection. 
Although our exploration on the dependence of the ejected mass on the stellar mass, the pulsation period, and other parameters are still incomplete and thus should be examined further in future studies, we use $M_\mathrm{ej}=10^{-3}\,\mathrm{M}_\odot$ as a typical ejected mass in a single envelope ejection episode. 
We note that the ejected mass may increase as the simulation progresses further, although our current simulations are truncated shortly after the appearance of the super-Eddington layer due to numerical issues. 

It is also important to specify the recurrence period of the mass ejection. 
One possibility is that the mass ejection repeatedly happens almost every pulsation cycles, once the star becomes unstable to radial pulsation. 
The models explored here exhibit pulsations with periods of $1$--$2$ yr in the linear stage. 
The pulsation periods appear to be a bit longer in the non-linear stage up to $\simeq 3$ yr. 
Therefore, one may assume that the recurrence time scale of a few years, $t_\mathrm{rec}\simeq 2$--$3$ yr. 
More conservatively, though, the envelope may have to wait a Kelvin-Helmholtz (KH) time for the oscillating part until it returns to the original state. 
Assuming that the mass of the oscillating part of the envelope is $M_\mathrm{env,osc}=$ a few $\mathrm{M}_\odot$ on top of the remaining inner matter with the mass given by $M_\mathrm{inner}\simeq 10\,\mathrm{M}_\odot$. 
The KH time is estimated to be 
\begin{eqnarray}
    t_\mathrm{kh}&=&\frac{GM_\mathrm{inner}M_\mathrm{env,osc}}{RL}
    \\
    &\simeq& 10
    \left(\frac{M_\mathrm{inner}}{10\,\mathrm{M}_\odot}\right)
    \left(\frac{M_\mathrm{env,osc}}{3\,\mathrm{M}_\odot}\right)  
    \left(\frac{R}{10^3R_\odot}\right)^{-1}
    \left(\frac{L}{10^5\mathrm{L}_\odot}\right)^{-1}\,\mathrm{yr}
    \nonumber
\end{eqnarray}
In this case, the recurrence time should be of the order of $t_\mathrm{rec}\simeq 10$ yr. 

The mass-loss rate averaged over a sufficient long time could then be 
\begin{equation}
    \dot{M}=\frac{M_\mathrm{ej}}{t_\mathrm{rec}}=10^{-4}
    \left(\frac{M_\mathrm{ej}}{10^{-3}\,\mathrm{M}_\odot}\right)
    \left(\frac{t_\mathrm{rec}}{10\,\mathrm{yr}}\right)^{-1}\,\mathrm{M}_\odot\,\mathrm{yr}^{-1}.
\end{equation}
This conservative estimate is higher than the mass-loss rate of RSGs with similar bolometric luminosity and surface temperature by a factor of a few -- several tens \citep[e,g,,][]{2014ARA&A..52..487S}. 
For example, the mass-loss formula by \cite{1988A&AS...72..259D} gives $\dot{M}\simeq 5.3\times 10^{-6}\, \mathrm{M}_\odot\,\mathrm{yr}^{-1}$ for $L/\mathrm{L}_\odot=10^5$ and $T_\mathrm{eff}=3500\,\mathrm{K}$. 
The formula by \cite{2005A&A...438..273V} is based on the compilation of dust-enshrouded RSGs and gives a higher value $\dot{M}\simeq 2.5\times 10^{-5}\, \mathrm{M}_\odot\,\mathrm{yr}^{-1}$ for the same $L$ and $T_\mathrm{eff}$. 
In addition, keeping this mass-loss rate entirely from the core C burning to the core-collapse ($\simeq 2\times 10^4\,\mathrm{yr}$), several $\mathrm{M}_\odot$ of the envelope could be lost. 
If pulsation-driven mass loss begins earlier in more massive stars, as we discussed above, such stars might shed their entire H-rich envelopes. 
In the next section, we consider possible impacts of the enhanced mass-loss to pre-SN evolution and CCSNe.

\section{Evolution beyond RSG}
In this section, we consider the evolution of massive stars beyond the RSG stage. 

\begin{figure*}
\begin{center}
\includegraphics[scale=0.9]{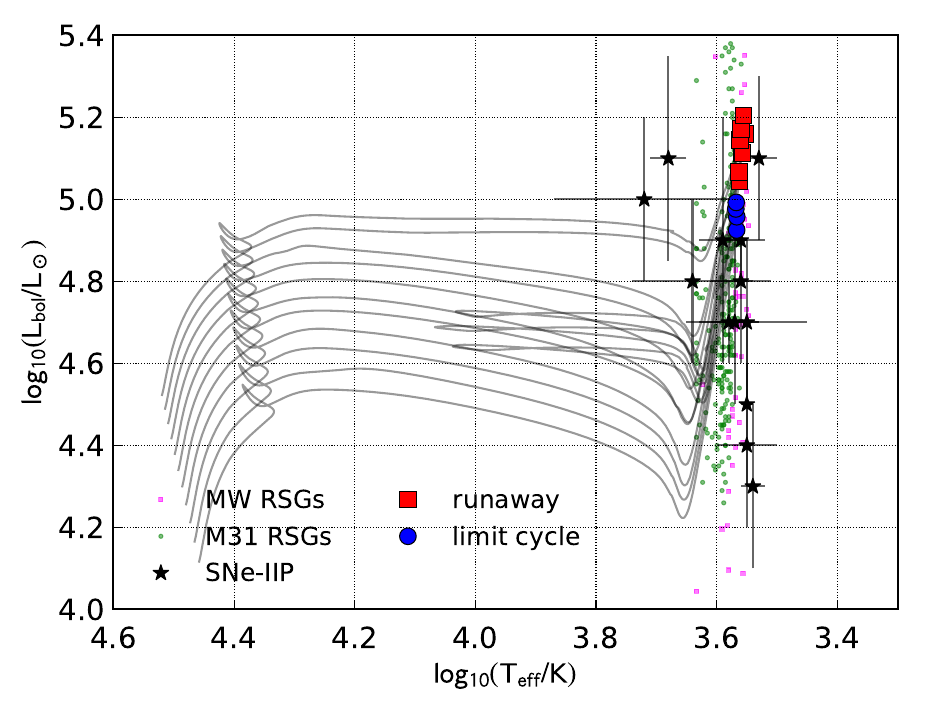}
\caption{Evolutionary tracks of our models. In each panel, the evolutionary tracks of the base models are plotted (thin gray line). 
When all the corresponding short-timestep models lead to limit cycle, the terminal point of the track is denoted by a filled circle (blue). 
On the other hand, filled squares (red) represent pulsation runaway cases. 
The SN-IIP progenitors \citep{2015PASA...32...16S} (black stars) and RSGs in Milky way (magenta squares) and M31 (green circles) are also presented. 
}
\label{fig:diagram}
\end{center}
\end{figure*}

We present the evolutionary tracks of our base models in Figure \ref{fig:diagram}. 
The models evolve along the standard track from the main-sequence, through the Hertzsprung gap, and to the red giant branch. 
In the following, we consider the locations of our evolutionary models in their final evolutionary states.

\subsection{Lower mass}
As we have clarified in previous sections, lower mass RSGs less likely experience the pulsation runaway. 
They are supposed to evolve as RSGs maintaining massive H-rich envelopes until the iron core collapse. 
The expected consequence of the evolution is therefore normal hydrogen-rich SNe. 

\subsubsection{Luminosity upper limit}
The models leading to limit cycles end up at locations below $\log(L/\mathrm{L}_\odot)\simeq 5.0$ in the HR diagram and are consistent with the domain occupied by the SNe-IIP progenitors ever identified \citep{2015PASA...32...16S}. 
Recently, \cite{2022MNRAS.515..897R} compared the luminosity distribution of RSGs in nearby galaxies and that of type II SN progenitors estimated in multiple ways, confirming an upper limit of $\log(L_\mathrm{bol}/\mathrm{L}_\odot)\simeq 5.1$ for SN-IIP progenitors. 
Therefore, our models leading to limit cycles are consistently in the luminosity range of SN-IIP progenitors. 
This agreement supports the idea that lower mass RSGs in our models explode as normal SNe-IIP.

As is pointed out above, however, the luminosity or mass threshold for the pulsation runaway would be dependent on parameters for convection. 
Also, RSGs around the boundary $L/M\simeq10^{3.9}\mathrm{L}_\odot/\mathrm{M}_\odot$ may eject only a small fraction the envelope during their limited lifetimes. 
Indeed, the models with $M_\mathrm{ini}=15.0$ and $15.5\,M_\odot$ are expected to experience pulsation runaway only after $t_\mathrm{cc}-t\simeq 2\times 10^3\,\mathrm{yr}$ (Tables \ref{table:model_description2} and \ref{table:model_description3}). 
Therefore, while it is true that luminous and massive RSGs are easier to cross the boundary than less massive stars, the region above $\log_{10}(L/\mathrm{L}_\odot)>5.1$ may not be a complete dessert for massive SN-IIP progenitors.  
Although its progenitor is unidentified, \cite{2018NatAs...2..574A} argue that SN~2015bs is likely a  type IIP SN arising from a massive RSG ($M_\mathrm{ini}=17$--$25\,\mathrm{M}_\odot$) in a low metallicity environment, $Z\sim 0.1Z_\odot$, based on its nebular spectra. 
Indeed, a metal-poor envelope less suffer from mass-loss, thereby making massive progenitors with lower $L/M$. 
As is suggested by \cite{2022A&A...660A..41M}, the paucity of SNe-IIP from massive RSGs could be regarded as the incompatibility with the initial mass function, rather than a sharp cut-off, which is also examined in more recent study by \cite{2024ApJ...969...57S}. 
For the assessment of this hypothesis, further dedicated efforts to identify SN progenitors are encouraged to reveal the luminosity function of SN-IIP progenitors.

\subsubsection{type II-P/L SNe}
The RSG models leading to limit cycles typically have the total masses of $12$--$13\,\mathrm{M}_\odot$ with remaining hydrogen-rich envelopes of $4$--$5\,\mathrm{M}_\odot$. 
\cite{2018PASA...35...49E} and \cite{2021ApJ...913...55H} present model grids of hydrodynamic and radiative transfer simulations of exploding RSGs with varied hydrogen-rich envelope masses. 
According to their simulation results, most of our models are expected to produce typical SN-IIP light curves with a $\sim 100\,\mathrm{days}$-long plateau, although some models with less massive envelope may fall into the type IIL (linearly declining) or short IIP (plateau duration of $\lesssim 70\,\mathrm{days}$) categories. 
The SN light curves are influenced by which phase, i.e., expansion or contraction, the envelope is at the moment of the iron core collapse \citep{2020ApJ...891...15G}. 
We note that the recent work by \cite{2025arXiv250811077B} and \cite{2025arXiv250811088L} also present diverse type II SN light curves depending on the phase of the pulsation.

\subsection{Higher mass}
For higher mass models, on the other hand, their envelopes lead to pulsation runaway, probably ejecting a part of the envelope. 
In the following, we consider the consequences of the radial pulsation and expected transients. 

\subsubsection{Envelope inflation or ejection?}
The \textsc{Mesa} computations with our current setups cannot tell exactly what happens when the outermost envelope expands faster than the surface escape velocity and the computation encounters boundary problems. 
The evolution of such an envelope including dynamical mass ejection would require dedicated mass reduction procedures as in \cite{clayton2018a}. 
As we have seen in the short-timestep models, the radial pulsation blocks a fraction of the radiation energy transported through the envelope and accumulates it in the form of the kinetic energy of the oscillating layers. 
In the following, we more generally consider how the envelope would respond to such an energy deposition in a short timescale. 

One of the possible responses of an RSG envelope to an energy injection is the envelope inflation. 
A number of previous studies on the evolution of massive stars (not limited to RSGs) with high luminosity-to-mass ratio have reported the so-called envelope inflation \citep{1999PASJ...51..417I,2006A&A...450..219P,2012A&A...538A..40G,2015A&A...573A..71K,2015A&A...580A..20S,2017A&A...597A..71S,2024arXiv241022403S}. 
It is a response of the stellar envelope against the radiation energy transport near the Eddington limit. 
As the name suggests, the stellar envelope expands to larger radii and layers with positive density and pressure gradients appear (density/pressure inversion). 
It is still debated whether or not such a state manifest itself \citep[e.g.,][]{2012A&A...537A.146E,2013MNRAS.433.1114Y}; it could be avoided by enhanced convective energy transport (i.e., larger $\alpha_\mathrm{mlt}$). 
Alternatively, several studies suggest that an optically thick wind occurs as a response \citep[e.g.,][]{1992ApJ...394..305K,2016ApJ...821..109R,2018ApJ...852..126N}. 
Even when such an inflated state is temporal, it would be possible that the iron core collapses and a CCSN happens in a star with an inflated radius, leaving unique electromagnetic signals \citep[e.g.,][]{2015A&A...575L..10M}. 

An alternative possibility is the superwind. 
The radial pulsations directly or indirectly (through shock thermalization) cause the swelling envelope.  
As is suggested for the mass-loss from asymptotic giant branch (AGB) stars \citep[e.g.,][and references therein]{2018A&ARv..26....1H}, the levitated atmosphere cools down to temperatures significantly below the original photospheric temperature of $\sim 3600$--$3800\,\mathrm{K}$. 
Dust grains can efficiently form in these cooled layers and then the layers with the enhanced opacity further absorb the stellar light and accelerate. 
In this scenario, the growing pulsation leads to an enhanced mass-loss and creates dusty circumstellar environments around massive pulsating RSGs. 
Dust formation are not taken into account in the present simulations. 
Therefore, its contributions to opacity is unclear. 

The envelope ejection may instead happen in a more eruptive way as we have seen in the model presented in Figure \ref{fig:ejection}. 
The enhanced mass-loss considered in Section \ref{sec:mass_loss} potentially remove a significant fraction of the hydrogen-rich envelope of RSGs in a dynamical way. 
There are several studies on the energy deposition in RSG envelopes in both instantaneous or continuous way \citep{2019ApJ...877...92O,2021MNRAS.500.1889O,2020A&A...635A.127K,2021A&A...646A.118K,2020ApJ...891L..32M,2022ApJ...930..168K,2022ApJ...936...28T}. 
These studies indeed demonstrate that the deposition of the energy comparable to the binding energy of the envelope leads to mass ejection in some cases. 
We note, however, that the way of converting the radiation energy into the kinetic energy of the oscillating layers can be different from that of the energy deposition explored in previous studies. 
While the energy injection at the core-envelope interface often results in lifting up the entire envelope against gravity rather than mass ejection, the pulsation could transmit the energy into the outermost oscillating layers. 
Indeed, we have seen an efficient transfer of the energy to growing pulsations rather than heating of the entire envelope in our simulations. 


\subsubsection{Enhanced mass-loss and post-RSG objects}
In the mass ejection scenario, it is also important to consider how the remaining star would look in its pre-SN stage. 
As we have seen above, the pulsations grow stronger for models with larger luminosity-to-mass ratios. 
Therefore, removing an outer part of the envelope makes the ratio even higher and thus the envelope can be even more likely to experience the pulsation-driven mass-loss when its radius is unchanged. 
The envelope may only be stabilized when the remaining envelope mass is significantly small ($<1\,\mathrm{M}_\odot$) and then the star turns into a blue/yellow supergiant with a smaller radius \citep{1998A&ARv...8..145D}. 
Several studies have already attempted an enhanced mass-loss in the RSG stage of massive stars and suggest that high-mass RSGs evolve back to bluer domain in the HR diagram \citep{2012A&A...537A.146E,2012A&A...542A..29G,2012A&A...538L...8G,2015A&A...575A..60M}. 
For example, \cite{2012A&A...537A.146E} and \cite{2012A&A...542A..29G} encounter unstable outer layers in their massive RSG models outshining nearly at the Eddington rates ($M_\mathrm{ini}>15\,\mathrm{M}_\odot$). 
Then, they assume RSG mass-loss rates artificially increased from their fiducial values by a factor of 3. 
As a result, their $20$ and $25\,\mathrm{M}_\odot$ models first evolve to RSGs and then turn to yellow supergiants before collapsing. 

Their evolutionary tracks are found consistent with the RSG population in M31 and M33 \citep{2023ApJ...942...69M}. 
Interestingly, such enhancements of the RSG mass-loss for luminous RSGs are also in line with the empirical $\dot{M}$--$L$ relations recently suggested by \cite{2023A&A...676A..84Y} and \cite{2024A&A...686A..88A},  who analyzed RSGs in small and large Magellanic clouds, respectively. 
Recently, \cite{2023A&A...678L...3V} suggest a parametrized mass-loss formula incorporating the dependence on the Eddington factor (a steeper function of $L$ for higher $L$) and demonstrate that it influences the upper mass limit of RSGs. 
They attribute the mass-loss enhancement to the multiple scattering effect in radiation-driven wind from a hot star. 
On the other hand, \cite{2024OJAp....7E..47F} consider the energy injection into the atmosphere of a star by shock waves excited by convection. 
They argue that such shock-supported chromosphere can account for the enhanced mass-loss. 
As we have seen above, the pulsation-driven mass ejection hypothesis can also lead to such a mass-loss enhancement.

\subsubsection{Luminous SNe-II}
Even in cases without mass ejection, observational outcome could be distinguished from normal SNe-IIP. 
\cite{2019ApJ...877...92O} have computed the response of the envelope of a $15\,\mathrm{M}_\odot$ RSG to continuous energy injection during the final several years prior to the iron core collapse. 
Their results suggest that the envelope is expanded beyond $\sim 10^{14}\,\mathrm{cm}$ rather than being ejected even with a super-Eddington energy injection rate $>10^{39}\,\mathrm{erg}\,\mathrm{s}^{-1}$. 
They find that the explosion of the star with the extended envelope results in a luminous hydrogen-rich SN, such as SN~2009kf \citep{2021MNRAS.500.1889O}.

\subsubsection{SNe-IIn and related transients}
It is also worth considering the expected transient populations when a non-negligible fraction of the H-rich envelope is lost in the late evolutionary stages. 
An enhanced mass-loss shortly before the core-collapse implies dense circumstellar environments around the SN progenitor. 
Indeed, a transient population showing strong narrow Balmer line emission, type IIn SNe \citep{1990MNRAS.244..269S,2012ApJ...744...10K,2014ARA&A..52..487S,2017hsn..book..403S}, suggests the presence of massive H-rich circumstellar material, presumably ejected shortly before the core-collapse. 
However, the mass-loss rate of $\sim 10^{-4}\,\mathrm{M}_\odot\,\mathrm{yr}^{-1}$ estimated in Section \ref{sec:mass_loss} only corresponds to the lower end of the mass-loss rates inferred from moderately luminous SNe-IIn \citep{2014ARA&A..52..487S}. 
Extreme events require even higher values, $0.01$--$1\,\mathrm{M}_\odot\,\mathrm{yr}^{-1}$. 
For such extreme cases, different mass-loss processes may play a dominant role, such as wave-driven mass ejection by gravity mode \citep{2012MNRAS.423L..92Q,2014ApJ...780...96S,2017MNRAS.470.1642F,2021ApJ...906....3W,2022ApJ...930..119W,2022ApJ...940L..27W}, as opposed to p-mode oscillation investigated in this work.

We note that the explored initial mass range of our model is narrow ($13$--$18\,\mathrm{M}_\odot$) and thus exploding RSGs with pulsation-driven mass ejection would not explain the whole type IIn SN population showing diverse properties \citep{2020A&A...637A..73N,2020ApJ...899...56S,2024arXiv241107287H}. 
Nevertheless, pulsation-driven mass ejection offers an interesting way to account for the production mechanism of the dense CSMs required for moderately luminous SNe-IIn.

\section{Final remarks and conclusion}\label{sec:summary}
In this work, we have investigated strong radial pulsations developing in massive RSGs in the late evolutionary stage as is previously demonstrated by several studies. 
Finally, we mention some caveats in our study and then summarize our findings below. 

First of all, our study relies on implicit hydrodynamic simulations implemented in \textsc{Mesa}. 
The development of radial pulsations of RSGs sharing similar properties with ours have been studied by different codes and settings \citep{1997A&A...327..224H,2010ApJ...717L..62Y}. 
These works and ours employ stellar evolutionary codes based on an implicit hydrodynamics solver, which introduces numerical viscosity and perhaps damping of oscillations. 
This may artificially influence the development of pulsations in the explored models. 
In particular, non-linear pulsations also involve supersonic flows, and shocks are expected to form.
However, the \textsc{Mesa} hydrodynamics module employing artificial viscosity is not optimal for shock-capturing. 
As a result, the non-linear behaviour may be sensitive to the parameters governing the strength of artificial viscosity (see also, discussion in \citealt{2020ApJ...902...63J}). 

Secondly, the stellar evolutionary models are still based on 1D spherical simulations. 
Although it is inevitable to rely on 1D simulations due to large dynamic ranges to resolve, the assumption of 1D spherical symmetry can exaggerate radially coherent motions. 
In realistic situation with no assumed symmetry, the convective regions in a star should have asymmetric 3D structure and excite non-radial oscillations. 
As a result, radial oscillations happen in more incoherent ways rather than efficiently increasing the amplitude of the fundamental mode. 
In order to resolve these issues, we ultimately need 3D hydrodynamic simulations of both the stellar core \citep[e.g.,][]{2007ApJ...667..448M,2009ApJ...690.1715A,2016ApJ...833..124M} and the envelope \citep[e.g.,][]{2012JCoPh.231..919F,2015ApJ...813...74J,2023Galax..11..105J,2024LRCA...10....2C} in advanced nuclear burning stages.

In relation to this issue, the treatment of the outermost layer warrants further discussion.
The envelope structures shown in Figures \ref{fig:limit_cycle} and \ref{fig:ejection} clearly distinguish between the inner convective region and the outer radiative layer, which plays a critical role in the development of super-Eddington layers in runaway cases. 
This clear separation at a narrow radial layer represents an averaged picture of an inherently three-dimensional atmospheric structure. 
In reality, the surface layers and the surroundings of luminous RSGs are characterized by large convective cells or blobs, with physical scales comparable to the stellar radius, as observed in spatially resolved surfaces of nearby RSGs \citep[e.g.,][]{2016A&A...588A.130M,2017Natur.548..310O,2018Natur.553..310P} and in their surroundings \citep[e.g.,][]{2001ApJ...551.1073H,2009A&A...504..115K,2011A&A...531A.117K}. 
Such large-scale convective motions extending beyond the stellar surface are also reproduced in global three-dimensional simulations of RSG atmospheres (e.g., \citealt{2010A&A...515A..12C,2022ApJ...929..156G,2024ApJ...962L..36M}; see \citealt{2023Galax..11..105J,2024LRCA...10....2C} for review).
Although a convective-to-radiative transition is indeed expected at the outermost layer, the non-spherical nature of the surface makes this transition region far more complex than the one-dimensional picture presented in this study. 
Therefore, accurate predictions of the properties of the pulsation-driven mass ejection, e.g., the ejected mass, ultimately require multi-dimensional simulations. 

Thirdly, growth of the pulsation and its behaviour in the non-linear stage appear to be coupled with convection and thus strongly depend on its numerical treatment. 
Adopting different numerical parameters in the time-dependent convection model can result in different velocity amplitudes realized during the non-linear stage. 
Consequently, the threshold mass and luminosity for pulsation runaway may also depend on the convection treatment and the choice of numerical viscosity, even though the qualitative scenario of ionization breakout and the subsequent acceleration of the super-Eddington layer remains valid. 
Therefore, further quantitative investigations into the growth and saturation of RSG pulsations are necessary.

Nevertheless, the strong pulsation reproduced in our models and possible pulsation-driven mass-loss can leave interesting observational fingerprints. 
Our findings are summarized as follows;
\begin{description}
    \item[(1)] 
Massive RSGs undergo strong radial pulsations in the late evolutionary stage.  
In the non-linear regime, the pulsations evolve into one of two distinct behaviours: limit cycle or pulsation runaway modes. 
The former is characterized by finite-amplitude oscillations in surface properties, with maximum expansion velocities remaining below the surface escape velocity. 
Consequently, such RSGs retain their massive H-rich envelopes until iron core collapse. 
In contrast, in the pulsation-runaway regime, pulsations grow strong enough to accelerate the outermost layers to velocities exceeding the escape velocity. 
\item[(2)]
The empirically obtained condition for avoiding pulsation runaway is governed by the luminosity-to-mass ratio, $L/M\lesssim 10^{3.9}\mathrm{L}_\odot/\mathrm{M}_\odot$ (Eq. \ref{eq:stability}). 
The corresponding luminosity upper limit is roughly consistent with that inferred for SN-IIP progenitors. 
However, the luminosity threshold would be dependent on the treatment of convection and its coupling to envelope oscillations. 
\item[(3)]
When the envelope undergoes pulsation runaway, a super-Eddington radiative layer develops, and part of the envelope may be expelled. 
As has been previously suggested \citep{1997A&A...327..224H,2010ApJ...717L..62Y,clayton2018a,2025arXiv250804497S}, such pulsation-driven mass-loss may contribute to the enhanced mass-loss rates required to reconcile stellar evolution models with observations of massive stars evolving toward the core-collapse. 
\end{description}

The findings presented above contribute to establishing a connection between massive stars within a specific mass range and various transient phenomena occurring before and after gravitational collapse; SN precursors, SNe-IIP, and SNe-IIn. 
Future pre- and post-explosion observations enabled by dedicated transient survey missions and monitoring programs targeting nearby massive stars, combined with continued advancements in theoretical modelling, will help uncover the final, and still largely mysterious, stages of massive star evolution leading up to core collapse. 

\section*{Acknowledgements}
The authors are grateful to the anonymous referee for his/her constructive comments that significantly improved the manuscript. 
The authors thank Jim Fuller for fruitful discussion during his visit to University of Tokyo. 
A.S. also thank Hideyuki Umeda and Ryosuke Hirai for fruitful discussion. 
A.S. acknowledges support by Japan Society for the Promotion of Science (JSPS) KAKENHI Grant Number JP22K03690. T.S. acknowledges support by JSPS KAKENHI Grant Numbers JP22K03688, JP22K03671, and JP20H05639.
The computations in this work are performed on Resceubbc PC cluster at the Research Center for the Early Universe, The University of Tokyo. 

\section*{Data Availability}
The inlist files and numerical data of our simulations are available upon request. 
 



\bibliographystyle{mnras}
\bibliography{refs} 




\appendix
\section{Resolution study}\label{sec:resolution_study}
\begin{figure*}
\begin{center}
\includegraphics[scale=0.70]{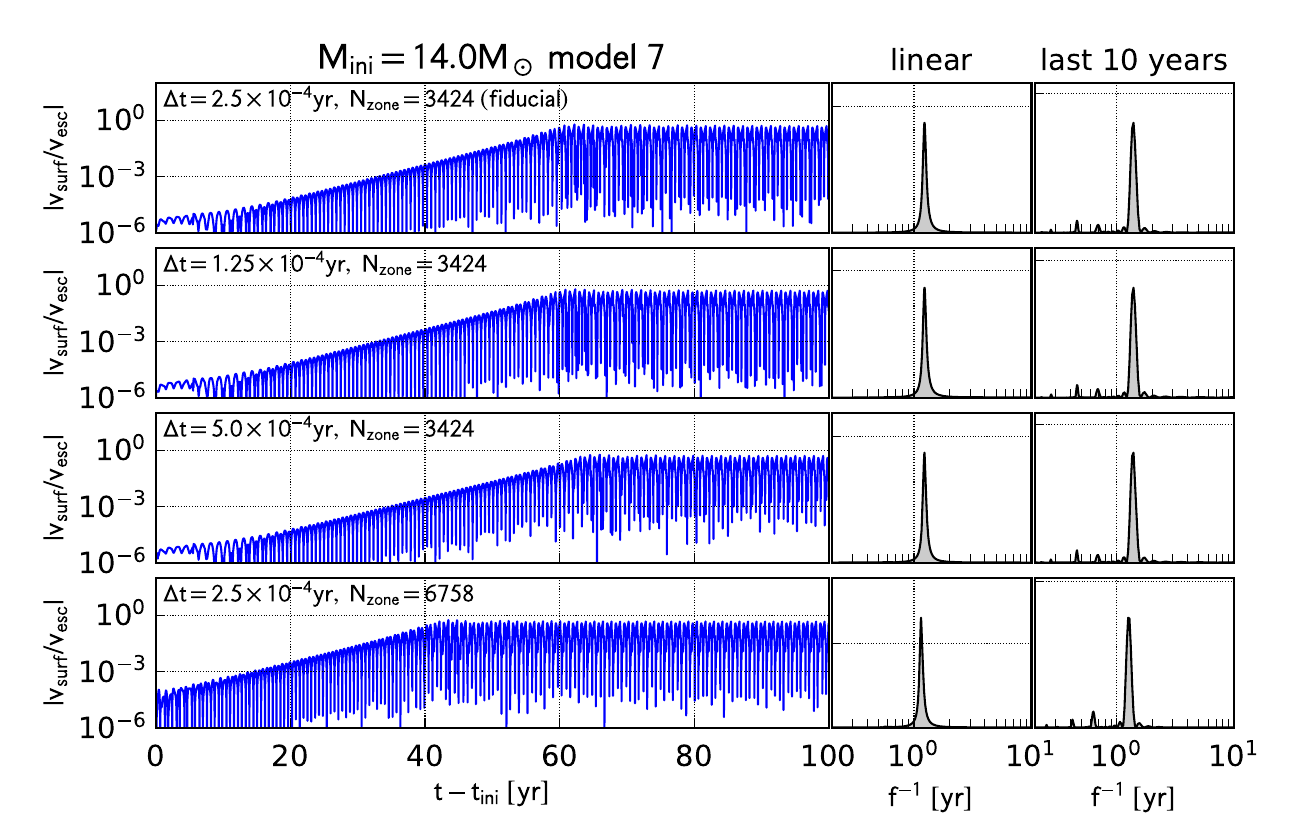}
\caption{Results of the models with different maximum timestep $\Delta t_\mathrm{max}$ and the number of zones $N_\mathrm{zone}$. 
The four rows show the evolution of the surface velocity and the results of the priod analyses presented in Figures \ref{fig:pulsation_M140} and \ref{fig:pulsation_M170}. 
The adopted $\Delta t$ and $N_\mathrm{zone}$ are presented in left panels. 
}
\label{fig:resolution_M140_pulse}
\end{center}
\end{figure*}
\begin{figure*}
\begin{center}
\includegraphics[scale=0.70]{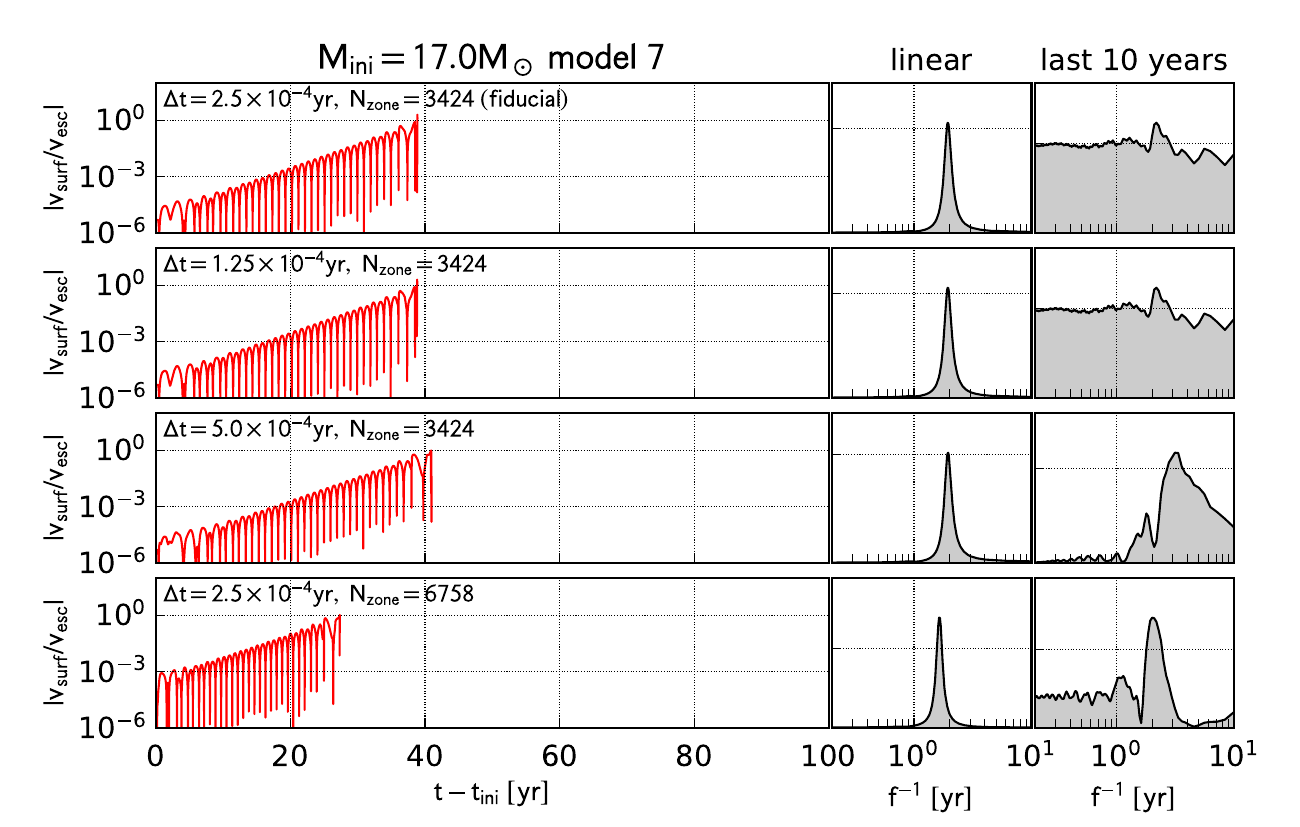}
\caption{Same as Figure \ref{fig:resolution_M140_pulse}, but for models with $M_\mathrm{ini}=17\,M_\odot$. 
}
\label{fig:resolution_M170_pulse}
\end{center}
\end{figure*}

\begin{figure*}
\begin{center}
\includegraphics[scale=0.68]{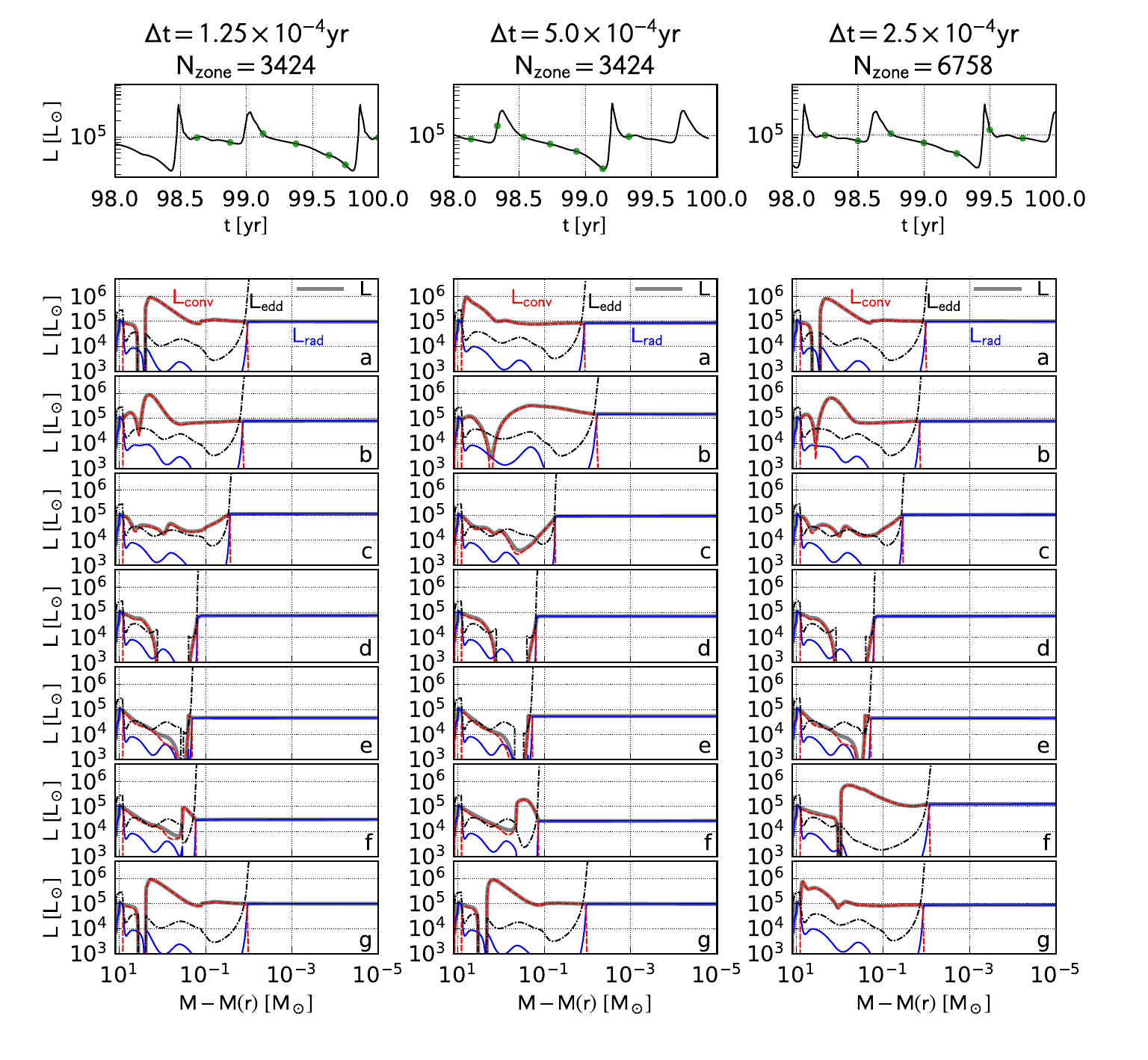}
\caption{The evolution of the envelope structure. 
The three columns corresponds to the models with different $\Delta t_\mathrm{max}$ and $N_\mathrm{zone}$ shown in Figure \ref{fig:resolution_M140_pulse}. 
In each column, the top panel represents the luminosity evolution during the last 10 years of the simulation, while the lower panels represent the luminosity as a function of the mass coordinate at the epochs marked in the top panel. 
}
\label{fig:resolution_M140_structure}
\end{center}
\end{figure*}
\begin{figure*}
\begin{center}
\includegraphics[scale=0.68]{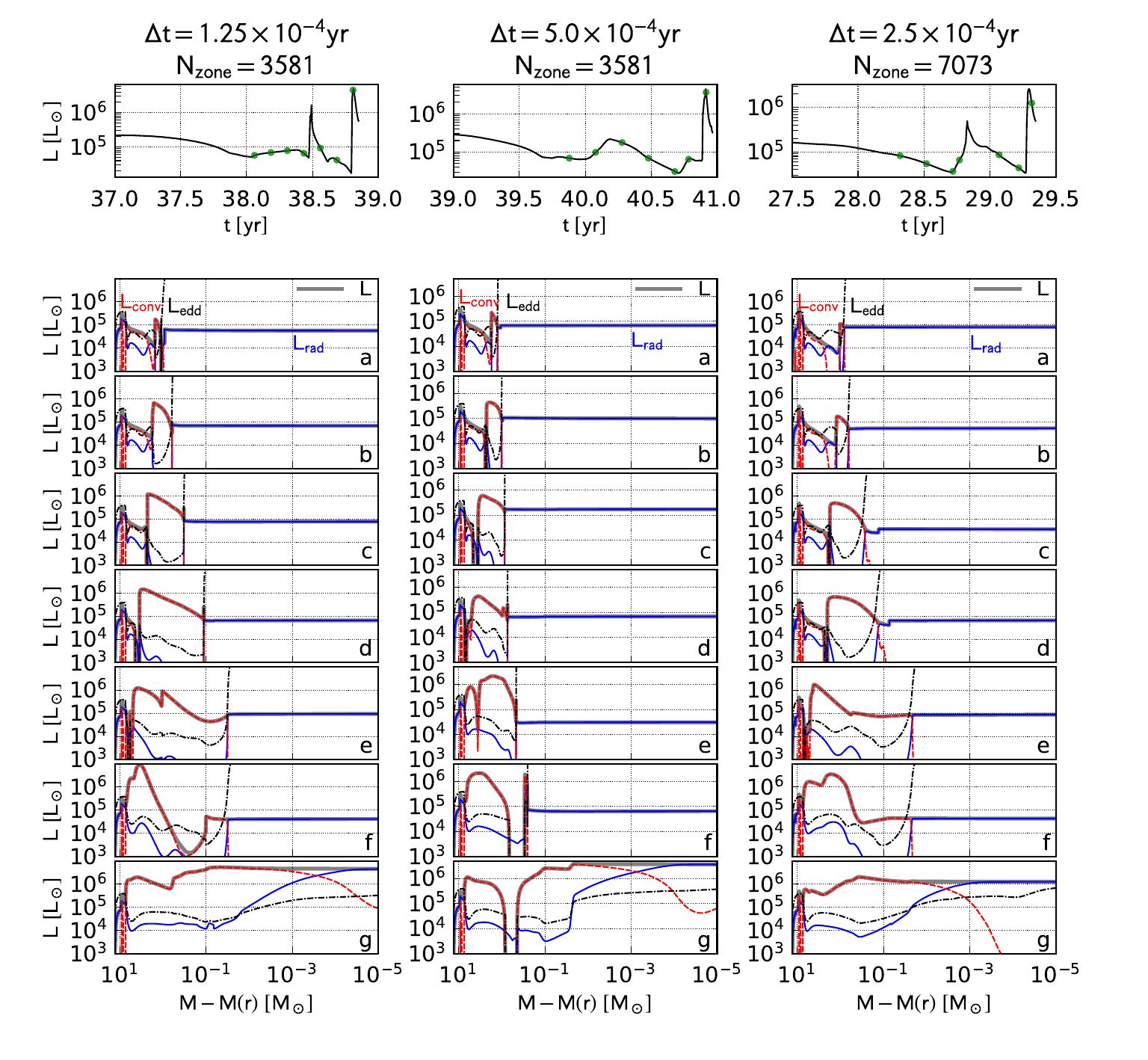}
\caption{Same as Figure \ref{fig:resolution_M140_structure}, but for the models with $M_\mathrm{ini}=17.0\,\mathrm{M}_\odot$. 
}
\label{fig:resolution_M170_structure}
\end{center}
\end{figure*}

In this section, we conduct simulations with different temporal and mesh resolutions to clarify how their results depend on the numerical setups. 

\subsection{Time step and mesh resolution}
We consider the dependence of pulsating behaviours on the time and mesh resolution by examining \texttt{M0140} and \texttt{M0170} model series (16 short-timestep models). 
In practice, we use the same initial snapshots as these short-timestep models, but instead of the fiducial time step limit of $\Delta t_\mathrm{max}=2.5\times 10^{-4}$ yr, we adopt the doubled and halved values $\Delta t_\mathrm{max}=5.0$ and $1.25\times 10^{-4}$ yr. 
For the influence of the mesh number, we adopt \texttt{delta\_mesh\_coefficient=0.25}, which roughly doubles the number of zones, for the same model set with the same maximum time step as the fiducial models. 

\subsection{Results}
We examine the $14\,\mathrm{M}_\odot$ and $17\,\mathrm{M}_\odot$ models approximately $10^3$ years prior to core collapse. 
In Figures \ref{fig:resolution_M140_pulse} and \ref{fig:resolution_M170_pulse}, we present the temporal evolution of the velocity amplitudes for these models. The maximum timestep values and the number of zones are indicated in the figures. 

The results for models with different timestep values closely match those of the fiducial models, suggesting that the adopted time resolution is sufficient to capture the correct pulsation mode and its growth. 
In contrast, models with higher mesh resolutions exhibit larger amplitudes in the beginning of the evolutions. 
Unlike modifications to timestep alone, increasing the mesh resolution requires restructuring the numerical grid. 
This process appears to introduce larger numerical perturbations, which seed the pulsation growth. 
Consequently, these models tend to experience either earlier saturation of the linearly growing pulsation or an earlier onset of pulsation runaway compared to the fiducial models. 
The corresponding periodograms for the higher-resolution models shown in Figures \ref{fig:resolution_M140_pulse} and \ref{fig:resolution_M170_pulse} display peaks at slightly shorter periods. 
Nevertheless, the differences in pulsation periods remain within approximately $10\%$, and the growth rates are consistent with those of the fiducial models. 
We also confirm that other short-timestep models with different time and mesh resolutions exhibit similar trends when compared to the fiducial models presented in Figures \ref{fig:pulsation_M140} and \ref{fig:pulsation_M170}.

In Figures \ref{fig:resolution_M140_structure} and \ref{fig:resolution_M170_structure}, we show the envelope structure for the same models analysed in Figures \ref{fig:resolution_M140_pulse} and \ref{fig:resolution_M170_pulse}. 
As in Figures \ref{fig:pulsation_M0140_model7} and \ref{fig:pulsation_M0170_model7}, we plot the luminosity as a function of the mass coordinate (measured from the surface) at seven characteristic epochs (labelled from {\it a} to {\it g}). 
The pulse profiles in the top rows of Figure \ref{fig:resolution_M140_structure} are similar across the models. 
The envelope snapshots shown in the lower panels reveal comparable structural features. 
In particular, the interface separating the convective and radiative layers is well resolved and evolves in a manner consistent with the fiducial model. 
This confirms that the non-linear pulsation behaviours discussed in Section \ref{sec:limit_cycle} are not artificially caused by either insufficiently short timesteps or inadequate mesh resolution. 

The snapshots in Figure \ref{fig:resolution_M170_structure} also show similar envelope structures and evolutions shortly before pulsation runaway in Figure \ref{fig:pulsation_M0170_model7}. 
The surface luminosity reaches a similar peak value, $L\simeq (3$--$5)\times 10^{6}\,\mathrm{L}_\odot$, associated with the release of the dissipated pulsation energy. 
Also, in all three models, the super-Eddington radiative layer appears after the luminosity maximum (phase {\it g}). 



\bsp	
\label{lastpage}
\end{document}